\documentclass[journal,12pt,draftclsnofoot,onecolumn]{IEEEtran}

\usepackage{cite}
\usepackage{epsfig}
\usepackage{epstopdf}
\usepackage{graphicx}
\usepackage{amsfonts}
\usepackage{amsmath}
\usepackage{amssymb}
\usepackage{multicol}
\usepackage{psfrag}
\usepackage{subfigure}
\usepackage{stfloats}
\usepackage{footnote}
\usepackage{array}
\usepackage{algorithm}
\usepackage{algorithmic}
\usepackage{color}
\usepackage{setspace}

\DeclareMathOperator*{\argmax}{argmax}

\newtheorem{theorem}{Theorem}
\allowdisplaybreaks

\begin{document}

\title{Hybrid Beamforming for Terahertz Multi-Carrier Systems over Frequency Selective Fading}

%\author{Hang Yuan,~\IEEEmembership{Student Member,~IEEE},
%        Nan Yang,~\IEEEmembership{Senior Member,~IEEE},\\
%        Kai Yang,~\IEEEmembership{Member,~IEEE},
%        Chong Han,~\IEEEmembership{Member,~IEEE},\\
%        and Jianping An,~\IEEEmembership{Member,~IEEE}
%        \vspace{-2em}
\author{Hang Yuan,
        Nan Yang,
        Kai Yang,
        Chong Han,
        and Jianping An
        %\vspace{-2em}

\thanks{Part of this work was accepted by {\em IEEE GLOBECOM 2018} \cite{yuan2018hybrid}.}
\thanks{H. Yuan, K. Yang, and J. An are with the School of Information and Electronics, Beijing Institute of Technology, Beijing 100081, China. H. Yuan is also with the Research School of Electrical, Energy and Materials Engineering, Australian National University, Canberra, ACT 2600, Australia (Email: \{yuanhang, yangkai, an\}@bit.edu.cn).}
\thanks{N. Yang is with the Research School of Electrical, Energy and Materials Engineering, Australian National University, Canberra, ACT 2600, Australia (Email: nan.yang@anu.edu.au).}
\thanks{C. Han is with the UM-SJTU Joint Institute, Shanghai Jiao Tong University, Shanghai 200240, China (e-mail: chong.han@sjtu.edu.cn).}}

\markboth{Submitted to IEEE Transactions on Communications}{}

\maketitle
\vspace{-4mm}

\begin{abstract}
We propose novel hybrid beamforming schemes for the terahertz (THz) wireless system where a multi-antenna base station (BS) communicates with a multi-antenna user over frequency selective fading. Here, we assume that the BS employs sub-connected hybrid beamforming and multi-carrier modulation to deliver ultra high data rate. We consider a three-dimensional wideband THz channel by incorporating the joint effect of molecular absorption, high sparsity, and multi-path fading, and consider the carrier frequency offset in multi-carrier systems. With this model, we first propose a two-stage wideband hybrid beamforming scheme which includes a beamsteering codebook searching algorithm for analog beamforming and a regularized channel inversion method for digital beamforming. We then propose a novel wideband hybrid beamforming scheme with two digital beamformers. In this scheme, an additional digital beamformer is developed to compensate for the performance loss caused by the constant-amplitude hardware constraints and the difference of channel matrices among subcarriers. Furthermore, we consider imperfect channel state information (CSI) and propose a probabilistic robust hybrid beamforming scheme to combat channel estimation errors. Numerical results demonstrate the benefits of our proposed schemes for the sake of practical implementation, especially considering its high spectral efficiency, low complexity, and robustness against imperfect CSI.
\end{abstract}


\IEEEpeerreviewmaketitle

\section{Introduction}\label{sec:intro}

Terahertz (THz) communications is an avant-garde wireless technology in the future beyond fifth generation (5G+) and sixth generation (6G) wireless networks \cite{Federici2010review}, due to its huge potential to support ultra-high-data-rate transmission \cite{akyildiz2014terahertz}, the Internet of nano-things \cite{akyildiz2010electromagnetic,akyildiz2014teranets}, and massive connectivity \cite{jornet2012phlame,an2020energy,gao2020contract} in the THz band (0.1--10 THz). Particularly, the drastically increasing traffic demand, which is expected to reach terabit per second (Tbps) within the next 5--10 years \cite{akyildiz2014terahertz}, can hardly be satisfied with the current wireless technologies that are designed in the microwave band below 6 gigahertz (GHz). The scarcity of available spectrum resources in the microwave band extremely restricts the achievable data rate, which has triggered tremendous research activities in higher-frequency bands \cite{Heathoverview,THzSubarrayMag}, such as the millimeter wave (mmWave) band ranging from 30 to 300 GHz. However, the data rate provided in the mmWave band is in the order of 10 gigabits per second (Gbps) \cite{rappaport2011state}, which is still below the expected traffic demand. Against this background, THz communications has become an indispensable enabler for Tbps links in future wireless data applications since it uses the ultra-large usable bandwidth which ranges from tens of GHz up to several THz \cite{akyildiz2014terahertz}. Importantly, the rapidly advancing THz device technologies, e.g., new graphene-based THz transceivers \cite{ju2011graphene} and ultra-broadband antennas that operate at THz frequencies \cite{liu2010broadband}, make THz communications a reality.

One critical factor that affects the propagation of THz waves is severe path loss, which may restrict THz transmission distance into a few meters. Thus, THz communication systems are very promising to be applied in an indoor environment \cite{Federici2010review}. The extremely short wavelength, however, has a superiority, i.e., a large number of antennas can be tightly packed into a small area at a transceiver. In \cite{han2018ultra}, an end-to-end channel model for ultra-massive MIMO communications in the THz band is developed by accounting for the properties of graphene-based plasmonic nano-antenna arrays. The developed model is then utilized to investigate the capacity of the ultra-massive MIMO system for both spatial multiplexing and beamforming regimes. Furthermore, spatial modulation techniques that leverage densely packed configurable antenna arrays are investigated to increase capacity and spectral efficiency \cite{Sarieddeen2019tera}. Hence, the high beamsteering gain, multiplexing gain, and spatial diversity gain enabled by massive multiple-input and multiple-output (MIMO) techniques can be exploited to combat severe path loss \cite{THzSubarrayMag}.

One challenge in realizing THz communications with massive MIMO is to deal with hardware constraints. By leveraging the unprecedented properties of nano-materials, such as graphene, plasmonic nano-antennas for THz communications have been proposed \cite{jornet2013graphene}. These nano-antennas are just tens of nanometers wide and few micrometers long, and can be integrated in nano-devices \cite{jornet2013graphene}. Moreover, new solutions with hybrid electronic-photonic approaches have been developed for practical THz transceiver designs \cite{Sengupta2018tera}. A significant advantage of using photonic components is that frequency-division multi-carrier systems can be easily realized by using the multi-wavelength laser source developed for optical multiplexing networks. However, components in radio frequency (RF) chains have much larger power consumption and higher complexity \cite{zhang2019mixed} compared to THz nano-antennas, imposing strict constraints on massive MIMO systems.

To tackle this challenge, a low-complexity indoor THz wireless system with hybrid beamforming was investigated in \cite{IndoorTHzComTWC}, where the number of RF chains is much less than the number of antennas.
In \cite{EEDesignTWC}, two hybrid beamforming architectures, namely, the fully-connected and sub-connected structures, were examined. Specifically, \cite{EEDesignTWC} showed that both the spectral efficiency and energy efficiency of the sub-connected structure are higher than those of the fully-connected structure under the consideration of insertion loss. Despite that \cite{IndoorTHzComTWC,EEDesignTWC} stand on their own merits, none of them has touched frequency selective fading in wideband THz systems. Since THz communications is anticipated to operate over broadband channels \cite{WidebandWaveform}, the design of frequency selective hybrid beamforming schemes is of great significance for wideband THz systems.

It is worthwhile noting that the research related to the design of frequency selective hybrid beamforming in THz systems is still in the infant stage. Some recent research efforts have been devoted to studying hybrid beamforming for mmWave systems over wideband channels.
With massive antennas in wideband mmWave systems, the transmit signals will be sensitive to the spatial-wideband effect, i.e., the physical propagation delay of electromagnetic waves traveling across the array becomes large \cite{wang2018spatial}. Considering the spatial-wideband effect, \cite{wang2019beam} proposed a channel estimation scheme for frequency-division duplex mmWave systems with hybrid beamforming architecture, where the frequency-insensitive parameters of each uplink channel path is extracted by a super-resolution compressed sensing approach.
In \cite{AlkhateebFS}, an optimal hybrid beamforming scheme that maximizes the achievable mutual information under total power and unitary power constraints was proposed for the orthogonal frequency-division multiplexing (OFDM)-based mmWave system with limited feedback channels.
A useful criterion for hybrid codebook construction was also developed in \cite{AlkhateebFS}, where both the baseband and RF precoders are taken from quantized codebooks.
Following the studies in \cite{AlkhateebFS}, \cite{ParkDS} considered the dynamic subarrays architecture for wideband hybrid beamforming, where a criterion for constructing the optimal subarrays that maximize the spectral efficiency was proposed. Furthermore, \cite{FSohrabiHybridBeamforming} showed that hybrid beamforming with a small number of RF chains can asymptotically approach the performance of fully digital beamforming for a sufficiently large number of transceiver antennas.
%which mainly arises from the sparse nature of mmWave channels.
The important findings in \cite{AlkhateebFS,ParkDS,FSohrabiHybridBeamforming} provide valuable insights and guidelines to the design of frequency selective hybrid beamforming, while having several limitations, as follows:
\begin{enumerate}
\item The molecular absorption effect which may hamper the signal propagation significantly in the mmWave and THz bands was not considered.
\item They did not exploit the high channel correlation among THz subcarriers, thus incurring a high computational complexity.
\item They assumed fully digital receiver and perfect carrier frequency offset (CFO) synchronization, which may be not practical. The CFO is a pervasive problem in THz communications, since the Doppler shift of THz channels is orders-of-magnitude larger than that of conventional microwave channels \cite{you2017BDMA}.
\item The robust design against imperfect channel state information (CSI) was not addressed. For massive MIMO systems, it is very difficult to obtain perfect CSI at the transmitter, especially for frequency selective channels.
\end{enumerate}
Due to the aforementioned limitations, the schemes developed in \cite{AlkhateebFS,ParkDS,FSohrabiHybridBeamforming} cannot be directly used in wideband THz wireless systems.

Against this background, we propose novel wideband hybrid beamforming schemes to enable ultra-high-data-rate transmission over THz frequency selective channels. In our considered system, a multi-antenna base station (BS) which employs the sub-connected architecture transmits to a multi-antenna user equipment (UE) using multi-carrier modulation. The rationale behind using multi-carrier modulation is that the very complex transceivers make OFDM very challenging to implement in THz wireless systems and spectrum resource is not scare in the THz band \cite{DABA}. Also, we focus on multiple subcarriers in one THz transmission window. For the sake of simplicity, we consider the equal power constraint for each subcarrier. The adaptive power constraint over subcarriers is an interesting direction, but not the focus of this paper. Our major contributions of this paper are summarized as follows.

\begin{itemize}
\item
We propose a wideband hybrid beamforming scheme taking into account the constraints on THz phase shifters. In this scheme, we develop a novel beamsteering codebook searching algorithm which aims to jointly maximize the sum of the normalized equivalent channel modulus over all subcarriers. The normalized factor which equals to the channel gain of each subcarrier can be estimated at the BS due to the unique dominant of THz line-of-sight (LOS) channels. We also develop a regularized channel inversion (RCI) method at the baseband with the power constraints. The rationale behind this scheme is to maximize the long-term average signal power for analog beamforming and minimize the inter-band interference (IBI) caused by the CFO for digital beamforming.
\item
We propose a novel wideband hybrid beamforming scheme including analog beamforming and two digital beamformings. Spatial channel covariance matrices are exploited to reduce the computational complexity in the design of analog beamforming. An additional digital beamforming is developed to compensate for the performance loss caused by two factors: 1) the constant-amplitude hardware constraints and
2) the difference of channel matrices among subcarriers. Along with the compensation digital beamforming, the statistical eigen hybrid beamforming can approximate the unconstrained optimal fully digital beamforming while requires much lower hardware complexity and power consumption.
\item
We consider imperfect CSI and further propose a robust wideband hybrid beamforming scheme. Instead of maximizing the SINR directly, our robust scheme maximizes the average signal power and introduces a probabilistic constraint to control the interference power, which ensures a very low probability that the interference power is higher than an acceptable level. The probabilistic constraint is transformed into a deterministic one, based on which a greedy amplitude-angle separate optimization method is developed.
\end{itemize}

In contrast to our preliminary work in [1], which only developed a beamsteering codebook searching algorithm for hybrid beamforming, this paper proposes two wideband hybrid beamforming schemes and relaxes the assumption that beamsteering codebooks are available at the BS. Also, [1] did not consider elevation angles, while this paper considers a three-dimensional (3D) propagation model with both azimuth and elevation angles. Furthermore, [1] considered perfect CSI, which might be difficult to obtain in massive MIMO THz systems. Against this drawback, this paper considers channel estimation errors and propose a robust scheme.

Aided by numerical results, we show that our proposed schemes achieve higher spectral efficiencies than the existing hybrid beamforming scheme proposed in \cite{EEDesignTWC}. We also show that the spectral efficiencies achieved by our proposed schemes are close to that achieved by the high-complexity fully digital beamforming scheme. We further show that the proposed robust hybrid beamforming scheme provides a clear performance advantage over the non-robust one in the presence of imperfect CSI. Overall, our numerical results demonstrate the benefits of our proposed schemes for practical implementation, especially considering its high spectral efficiency, low complexity, and robustness.

\emph{Notation:} Scalar variables are denoted by italic symbols. Vectors and matrices are denoted by lower-case and upper-case boldface symbols, respectively. Given a vector $\mathbf{z}$, $(\mathbf{z})_{i}$ denotes the $i$th entry in $\mathbf{z}$ and $\measuredangle\left(\mathbf{z}\right)$ denotes the angle of $\mathbf{z}$. Given a matrix $\mathbf{Z}$, $\|\mathbf{Z}\|_\emph{F}$ denotes the Frobenius norm of $\mathbf{Z}$, $\mathbf{Z}^{-1}$ denotes the inverse of $\mathbf{Z}$, $\mathbf{Z}^{T}$ denotes the transpose of $\mathbf{Z}$, $\mathbf{Z}^{H}$ denotes the Hermitian transpose of $\mathbf{Z}$, ${\textrm{tr}}\left\{\mathbf{Z}\right\}$ denotes the trace of $\mathbf{Z}$, $\textrm{eig}_{1}\left(\mathbf{Z}\right)$ denotes the first eigenvector of $\mathbf{Z}$ corresponding to the largest eigenvalue, and $[\mathbf{Z}]_{r,:}$ and $[\mathbf{Z}]_{:,c}$ denote the $r$th row and $c$th column of $\mathbf{Z}$, respectively. Furthermore, $\mathbf{I}_{m}$ denotes the $m\times{m}$ identity matrix, ${\textrm{diag}}\{\cdot\}$ denotes a diagonal matrix with indicated vector along the diagonal, and $\mathcal{CN}\left(\mu,\nu\right)$ denotes the complex Gaussian distribution with mean $\mu$ and covariance $\nu$,
and $\mathbb{E}\left[\cdot\right]$ denotes the expectation.

\section{System and Channel Models}\label{sec:model}

\subsection{System Model}\label{sec:system_model}

\begin{figure*}[!t]
    \begin{center}
        \includegraphics[width=0.7\columnwidth]{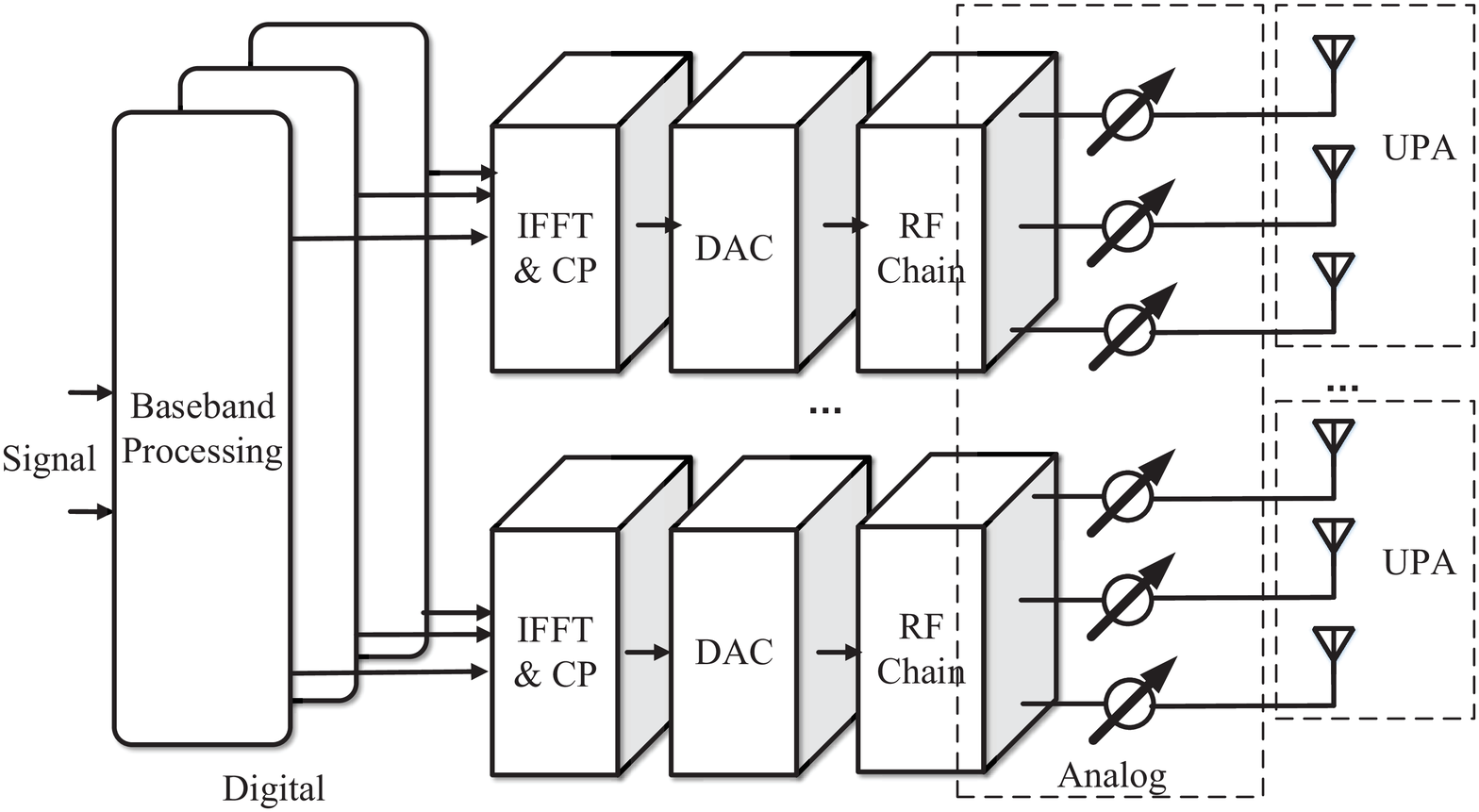}
        \caption{Illustration of the sub-connected hybrid beamforming architecture at the BS.}
        \label{Fig_System_Model}
    \end{center}\vspace{-4mm}
\end{figure*}

We consider an indoor THz wireless communication system where an $N_{\textrm{BS}}$-antenna BS transmits to an $N_{\textrm{U}}$-antenna UE using the ultra-large THz band.
The BS employs the sub-connected hybrid beamforming architecture to deploy a large number of antennas via $N_{\textrm{RF}}$ RF chains, as depicted in Fig. \ref{Fig_System_Model}. With this architecture, each RF chain drives one disjoint subarray only. We assume that each subarray at the BS is a uniform planar array (UPA) with $M_{t}\times{}N_{t}$ tightly-packed directional antennas, each of which is attached to a THz phase shifter \cite{IndoorTHzComTWC}. Such THz phase shifters are implemented by integrally-gated graphene transmission-lines \cite{chen2013tera} with small return loss, insertion loss, and phase error. The total number of antennas at the BS is $N_{\textrm{BS}}=N_{\textrm{RF}}M_{t}N_{t}$.
At the UE, only one subarray driven by a single RF chain is equipped due to UE's constraints on hardware and signal processing complexity \cite{EEDesignTWC}. The subarray at the UE is a UPA with $M_{r}\times{}N_{r}$ tightly-packed antennas.
The total number of antennas at the UE is $N_{\textrm{U}}=M_{r}N_{r}$.

%We assume that the antennas within each subarray are equally spaced, where the space between two adjacent antennas, denoted by $a$, is smaller than the wavelength. Generally, $a$ is in the order of half wavelength to avoid the grating lobe effect in beamforming \cite{alice2019ultra}.
%We also assume that the space between two adjacent subarrays, denoted by $b$, is larger than the wavelength, which indicates that beamsteering directions among different subarrays can be designed individually with the sub-connected architecture. In this case, all signal processing is conducted at the subarray level. Thus, the large channel matrix between the BS and UE can be decomposed into $N_{\textrm{RF}}$ separate sub-matrices.

The BS adopts the multi-carrier modulation to transmit signals. We assume that the data transmitted to the UE is modulated and spread across $K$ non-overlapping subcarriers. The subcarrier spacing is assumed to be equal to the subcarrier bandwidth. %Therefore, the BS transmits a length-$K$ data symbol to the UE during one block.
At the BS, each subcarrier sends one data stream. The transmit data, $s[k]$, is constrained by the transmit power given by $\mathbb{E}\left[s[k]s[k]^{*}\right] =\frac{P_s}{K}$, where $P_s$ is the total transmit power over all subcarriers.
At the baseband, digital beamforming is employed to control and route data streams through different RF chains. For example, $s[k]$ is precoded by an $N_\text{RF} \times 1$ digital baseband beamformer $\mathbf{f}_{\textrm{BB}}[k]$ in the frequency domain. Then, the precoded signal is transformed into the time domain using $N_\text{RF}$ parallel $K$-point inverse fast Fourier transform (IFFT). After this, a length-$Q$ cyclic prefix is added to the signal on each RF chain to eliminate the inter-symbol interference, before applying the $N_{\textrm{BS}}\times{}N_{\textrm{RF}}$ analog RF beamformer $\mathbf{W}$ in the time domain. Since the RF beamforming is implemented after the IFFT, the analog RF beamformer is consistent for all subcarriers. Hence, the transmitted symbol at the $k$th subcarrier is expressed as $\mathbf{x}[k]=\mathbf{W}\mathbf{f}_{\textrm{BB}}[k]s[k]$.

For the sub-connected architecture, the analog RF beamformer $\mathbf{W}$ is a block diagonal matrix which can be expressed as $\mathbf{W}=\textrm{diag}\{\mathbf{w}_{1}, \mathbf{w}_{2},\ldots,\mathbf{w}_{N_\text{RF}}\}$,
%\begin{align} \label{eq sec2 W}
%\mathbf{W}=\begin{bmatrix}
%\mathbf{w}_{1} &\mathbf{0} & \cdots &\mathbf{0}\\
%\mathbf{0} & \mathbf{w}_{2} & \cdots & \mathbf{0} \\
%\vdots & \vdots & \vdots & \vdots \\
%\mathbf{0} & \mathbf{0} & \cdots & \mathbf{w}_{N_\text{RF}}
%\end{bmatrix},
%\end{align}
where $\mathbf{w}_{n}$, $n\in\{1, 2, \ldots, N_{\textrm{RF}}\}$, is an $M_{t}N_{t}\times1$ vector. Each entry in $\mathbf{w}_{n}$ is limited by a constant modulus constraint such that $\left|(\mathbf{w}_{n})_{i}\right|=1/\sqrt{M_{t}N_{t}}$, $i\in\{1,2,\ldots,M_{t}N_{t}\}$. Furthermore, we consider the equal power constraint for each subcarrier, which is imposed by normalizing $\mathbf{F}_{\textrm{BB}}[k]$ such that $\|\mathbf{W}\mathbf{f}_{\textrm{BB}}[k]\|^{2}_{F}=1$.

At the UE, the received signal at the $k$th subcarrier is given by $y[k]=\mathbf{v}^{H}\mathbf{H}[k]\mathbf{x}[k]+n[k]$,
%\begin{align} \label{eq sec2 y_u}
%y_{k}=\mathbf{v}^{H}\mathbf{H}_{k}\mathbf{x}_{k}+n_{k},
%\end{align}
where $\mathbf{v}$ is the $N_{U}\times1$ receive analog beamforming vector, $\mathbf{H}[k]$ is the $N_{U}\times N_{\textrm{BS}}$ THz channel matrix between the BS and the UE, and $n[k]$ denotes the additive Gaussian white noise (AWGN) at the UE with the power of $\sigma_{n}^{2}$. We note that $\mathbf{v}$ is also limited by the constant modulus constraint such that $\left|(\mathbf{v})_{j}\right|=1/\sqrt{N_{U}}$, $j\in\{1,2,\ldots,N_{U}\}$.

\subsection{Frequency Selective THz Channel Model}\label{sec:channel_model}

At very high frequencies, the molecular absorption loss drastically affects the propagation. The molecular absorption loss varies across frequencies and creates many high attenuation peaks which define multiple transmission windows \cite{THzSubarrayMag,ChannelModelTWC}. The bandwidth of such windows highly depends on the humidity and the transmission distance, and can easily exceed $100$ GHz. In this work, we focus on the first absorption-defined transmission window in the THz band, i.e., the carrier frequency is set within $300 - 450$ GHz.

We note that each transmission window experiences highly frequency selective fading. Indeed, the bandwidth of the transmission window, which is approximately $0.2$ THz \cite{THzSubarrayMag}, is much larger than the coherence bandwidth, which is approximately $1$ GHz in the indoor environment \cite{WidebandWaveform,MultirayWidebandCharact}. The unique frequency selectivity in the THz band comes from the multi-path effect, as well as the molecular absorption. This is because the molecular absorption loss is not constant across frequencies, but rather highly frequency-dependent. The absorption coefficients for different molecules varies significantly with the transmission frequency. The level of frequency-selectivity in the THz channel increases with center frequency, pulse bandwidth, and communication distance \cite{WidebandWaveform}.

Due to the high frequency dependence in THz channels, we build up the channel model as the combination of many subcarriers. In each subcarrier, the arrival paths generally consist of a LOS ray and several non-LOS (NLOS) rays that are caused by reflection. %The scattering and diffraction can be neglected since they impose a negligible effect on propagation in the THz band.
Thus, the delay-$q$ channel response matrix is given by \cite{StatisticalModelJSAC,StocModelTWC}
\begin{align}\label{eq sec2 H_uq}
\mathbf{H}_{q}(f_k,d)%=\mathbf{H}^{\textrm{L}}_q(f,d)+\mathbf{H}^{\textrm{NL}}_q(f,d) \notag\\
=&\alpha^{\textrm{L}}(f_k,d) p_r\left(qT_s-\tau\right)G_{t}G_{r}\mathbf{a}_r\left(\theta^{r},\phi^{r}\right) \mathbf{a}^{H}_t\left(\theta^{t},\phi^{t}\right)\notag\\
+&\sum_{i=1}^{N_\textrm{clu}}\sum_{l=1}^{L^{i}_\textrm{ray}}\alpha^{\textrm{NL}}_{i,l}(f_k,d)p_r\left(qT_s-\tau_{i,l}\right)
G_{t}G_{r}\mathbf{a}_r\left(\theta^{r}_{i,l},\phi^{r}_{i,l}\right)
\mathbf{a}^{H}_t\left(\theta^{t}_{i,l},\phi^{t}_{i,l}\right),
\end{align}
where $f_k$ is the center frequency of the $k$th subcarrier, $d$ is the transmission distance, $N_\textrm{clu}$ is the number of clusters, $L^{i}_\textrm{ray}$ is the number of rays in the $i$th cluster, $\alpha^{\textrm{L}}$ and $\alpha^{\textrm{NL}}_{i,l}$ are the complex gains of the LOS ray component and the NLOS ray component, respectively, $G_{t}$ and $G_{r}$ are the associated transmit and receive antenna gains, respectively, and $p_r\left(qT_{s}-\tau\right)$ is the pulse-shaping function for $T_{s}$ space signaling at the time delay of $\tau$ \cite{AlkhateebFS}. Moreover, we denote $\mathbf{a}_{t}\left(\cdot,\cdot\right)$ and $\mathbf{a}_{r}\left(\cdot,\cdot\right)$ as the antenna array response vectors at the transmitter and receiver, respectively, where $\theta^{t}$, $\phi^{t}$, $\theta^{r}$, and $\phi^{r}$ denote the azimuth AoD (A-AoD), elevation AoD (E-AoD), azimuth AoA (A-AoA), and elevation AoA (E-AoA) for the LOS ray, respectively, and $\theta^{t}_{i,l}$, $\phi^{t}_{i,l}$, $\theta^{r}_{i,l}$, and $\phi^{r}_{i,l}$ denote the A-AoD, E-AoD, A-AoA, and E-AoA for the $l$th NLOS ray in the $i$th cluster, respectively.

Next, we give a brief introduction to some unique characteristics of THz channels that are different from those of mmWave channels. Such characteristics play a fundamental role in our design and analysis.

\subsubsection{Path Gain}

The path loss in THz channels consists of spreading loss and molecular absorption loss, which is given by $\left|\alpha^{\textrm{L}}\right|^{2}=L_\textrm{spr}\left(f,d\right)L_\textrm{abs}\left(f,d\right) =\left(\frac{c}{4\pi{}fd}\right)^{2}e^{-k_\textrm{abs}(f)d}$,
with $c$ being the speed of light and $k_\textrm{abs}(f)$ being the frequency-dependent medium absorption coefficient. The absorption coefficient is determined by the composition of the transmission medium at the molecular level \cite{ChannelModelTWC}. Compared to mmWave channels, both spreading loss and molecular absorption loss become much more severe in THz channels. Also, the major cause of the molecular absorption in THz channels comes from water vapor, which indicates that the molecular absorption loss is humidity-dependent.

Due to the high reflection loss in THz channels, we note that THz channels only have a few NLOS paths and are much sparser than mmWave channels. The gap between the LOS and NLOS path gains in THz channels (e.g., more than $15$ dB on average \cite{han2018ultra}) is more significant than that in mmWave channels. Hence, THz channels are apparently LOS-dominant and NLOS-assisted, and more sensitive to blockages than mmWave channels. This property also motivates us to build a 3D THz channel model since THz BSs are very likely to be deployed at a high place to prevent blockages.

\subsubsection{Angles of Departure and Arrival}

The angular spread of the THz channel in indoor environments is much smaller than that in
mmWave or microwave channels, due to the significant increase in reflection and scattering loss and the decrease in the number of NLOS rays \cite{THzSubarrayMag}. We clarify that $\theta^{t}_{i,l}$ and $\phi^{t}_{i,l}$ in the same cluster consist of the mean AoD and the AoD shift for the $l$th NLOS ray, while $\theta^{r}_{i,l}$ and $\phi^{r}_{i,l}$ consist of the mean AoA and the AoA shift for the $l$th NLOS ray. Thus, $\theta^{t}_{i,l}$, $\phi^{t}_{i,l}$, $\theta^{r}_{i,l}$, and $\phi^{r}_{i,l}$ are expressed as $\theta^{t}_{i,l}=\Theta^{t}_i+\vartheta^{t}_{i,l}$, $\phi^{t}_{i,l}=\Phi^{t}_i+\varphi^{t}_{i,l}$,
$\theta^{r}_{i,l}=\Theta^{r}_i+\vartheta^{r}_{i,l}$, and $\phi^{r}_{i,l}=\Phi^{r}_i+\varphi^{r}_{i,l}$,
respectively, where $\Theta^{t}_i/\Phi^{t}_i$ and $\Theta^{r}_i/\Phi^{r}_i$ denote mean AoD and mean AoA of the $i$th cluster, and $\vartheta^{t}_{i,l}$, $\varphi^{t}_{i,l}$, $\vartheta^{r}_{i,l}$, and $\varphi^{r}_{i,l}$ denote angle shifts. We note that $\Theta^{t}_i/\Phi^{t}_i$ and $\Theta^{r}_i/\Phi^{r}_i$ follow uniform distributions on $[-\pi,\pi]$ and $[-\frac{\pi}{2},\frac{\pi}{2}]$, respectively.
We also note that $\vartheta^{t}_{i,l}$, $\varphi^{t}_{i,l}$, $\vartheta^{r}_{i,l}$, and $\varphi^{r}_{i,l}$ can be characterized by a zero-mean second order Gaussian mixture model (GMM)~\cite{AoAAoDToAConf}.

%\subsubsection{Antenna Gains}
%
%In THz communications, highly directional antennas and high antenna gains are inevitable due to the high path loss. To characterize the antenna radiation pattern, we model the transmit antenna gain $G_t$ for a single beam with beam width of $\varphi$, which determines the antenna directionality. The transmit antenna gain $G_t$ for the main lobe is given by
%\begin{align}
%G_t=\frac{2}{1-\cos(\varphi/2)}.
%\end{align}
%We note that for $\varphi=2\pi$, i.e., an ideal omnidirectional antenna, the transmit antenna gain $G_t=1$.

Given the delay-$q$ THz channel model in \eqref{eq sec2 H_uq}, the channel at the $k$th subcarrier is finally expressed as
\begin{align} \label{eq sec2 H_uk}
\mathbf{H}[k]&=\sum_{q=1}^{Q}\mathbf{H}_{q}(f_k,d)e^{-j\frac{2\pi k}{K}q}\notag\\
&=\alpha^{\textrm{L}}(f_k,d) G_t G_r \mathbf{a}_r\left(\theta^{r},\phi^{r}\right) \mathbf{a}^{H}_t\left(\theta^{t},\phi^{t}\right) P_r(k,\tau)\notag\\
&+\sum_{i=1}^{N_\textrm{clu}}\sum_{l=1}^{L^{i}_\textrm{ray}}\alpha^{\textrm{NL}}_{i,l}(f_k,d) G_t G_r \mathbf{a}_r(\theta^{r}_l,\phi^{r}_l) \mathbf{a}_r\left(\theta^{r}_{i,l},\phi^{r}_{i,l}\right)
\mathbf{a}^{H}_t\left(\theta^{t}_{i,l},\phi^{t}_{i,l}\right),
\end{align}
where $P_r(k,\tau)$ is defined as $P_r(k,\tau)=\sum_{q=1}^{Q}{p_r\left(qT_s-\tau\right)e^{-j\frac{2\pi k}{K}q}}$.
We clarify that all subcarriers are low-rank channels. Indeed, $\mathbf{H}[k]$ contains a very limited number of NLOS rays. We also clarify that the channel coefficients across subcarriers are highly correlated since $\mathbf{a}_r(\theta^{r},\phi^{r})$ and $\mathbf{a}_t(\theta^{t},\phi^{t})$ are the same for all subcarriers.

So far, we have established the system and channel models. Such models enable us to formulate the hybrid beamforming design problem in the next subsection.

\subsection{Problem Formulation of Hybrid Beamforming Design}\label{sec:design_formulation}

In multi-carrier systems, the accurate frequency synchronization operating in the THz band is extremely challenging since it requires very high sampling rates (e.g., over multi-giga- or tera-samples per second) \cite{MultirayWidebandCharact}. Hence, we consider the CFO of multi-carrier systems in this paper for the sake of practicality. The CFO may be caused by misalignment in carrier frequencies or Doppler shift. The Doppler shift of THz channels is orders-of-magnitude larger than that of conventional microwave channels \cite{you2017BDMA}. The CFO of THz channels can be estimated by a under-sampling approach with narrow-band filtering and coprime sampling \cite{song2018frequency}.
The combined signal at the $k$th subcarrier at the UE can be written as \cite{SINRinSVchannelsJSAC}
\begin{align}\label{eq sec3 y_u}
\tilde{y}[k]=S_0\mathbf{v}^{H}\mathbf{H}[k]\mathbf{W}\mathbf{f}_{\textrm{BB}}[k]s[k]
+\Delta+n[k],
\end{align}
where $\Delta=\sum_{\lambda=1,\lambda\neq{}k}^{K}S_{\lambda-k}\mathbf{v}^{H}\mathbf{H}[\lambda]\mathbf{W}\mathbf{f}_{\textrm{BB}}[k]s[k]$
denotes the IBI caused by other subcarriers. The sequence $S_i$, $i\in\{1\!-\!K,\ldots,0,\ldots,K\!-\!1\}$, denotes the IBI coefficient which depends on the CFO and is given by \cite{{PewithCFO}}
\begin{align}\label{eq sec3 S_i}
S_i=\frac{\sin\pi(i+\varepsilon)}{K\sin\frac{\pi}{K}(i+\varepsilon)} e^{j\pi\left(1-\frac{1}{K}\right)(i+\varepsilon)},
\end{align}
where $\varepsilon$ is the ratio between the CFO and the subcarrier spacing.
For zero CFO, $S_i$ reduces to the unit impulse sequence. We clarify that IBI mainly captures the power leakage from neighboring subcarriers.
Thus, the IBI decreases when the bandwidth of subcarriers increases.
%When the transmit distance increases, the frequency selectivity of THz channels becomes more severe, leading to a decrease in the IBI. However,
With the employment of directional antennas, the path loss and delay spread of channels are reduced and the IBI also increases \cite{PewithCFO}.

Given the received signal in \eqref{eq sec3 y_u}, the average achievable data rate is derived as
\begin{align}\label{eq sec3 Rate}
R&=\frac{1}{K}\sum_{k=1}^{K}B\log_2\left(1+\gamma_k\right)\notag\\
&=\frac{1}{K}\sum_{k=1}^{K}B\log_2\left(1\!+\!\frac{|S_{0}|^2\left|\mathbf{v}^{H}\mathbf{H}[k] \mathbf{W}\mathbf{f}_{\textrm{BB}}[k]\right|^2}{\sum_{\lambda=1,\lambda\neq{}k}^{K}|S_{\lambda-k}|^{2}
\left|\mathbf{v}^{H}\mathbf{H}[\lambda] \mathbf{W}\mathbf{f}_{\textrm{BB}}[k]\right|^{2}\!+\!\psi}\right),
\end{align}
where $\gamma_k$ is the SINR at the $k$th subcarrier, $B$ denotes the bandwidth of a subcarrier, and $\psi=K\sigma^2_{n}/P_{s}$. Adopting $R$ as the system performance metric, the hybrid beamforming design problem is to find $\mathbf{v}$, $\{\mathbf{w}_n\}^{N_\text{RF}}_{n=1}$, and $\{\mathbf{f}_{\textrm{BB}}[k]\}^{K}_{k=1}$ such that $R$ is maximized.
Due to the constraints on RF hardware, the analog beamforming vectors generally take certain values. Hence, in the next section we will investigate the hybrid beamforming problem when analog beamformers are taken from quantized codebooks while no quantization constraints are imposed on digital beamformers.

\section{Wideband Hybrid Beamforming Scheme with Beamsteering Codebooks}\label{sec:design_codebook}

In this section, we propose a two-stage wideband codebook-based hybrid beamforming scheme for the considered system. %based on both the THz channel characteristics and the hardware constraints described in Section~\ref{sec:model}.
In the first stage, we develop a novel beamsteering codebook searching algorithm for analog beamforming. Compared to the narrow-band beamsteering codebook searching algorithm in \cite{IndoorTHzComTWC,EEDesignTWC}, we propose to jointly maximize the sum of normalized equivalent channel modulus over all subcarriers, to obtain the optimal analog beamformer serving all subcarriers. Here, we develop an unique normalized factor which is equal to the LOS path gain to ensure the fairness among subcarriers.
%In the first stage, we develop a novel beamsteering codebook searching algorithm for analog beamforming. Here, we propose to jointly maximize the sum of normalized equivalent channel modulus over all subcarriers to ensure the fairness among subcarriers.
In the second stage, we design a revised RCI method with IBI elimination for digital beamforming. %This guarantees the low complexity of our proposed hybrid beamforming scheme.

In our considered system, analog beamforming is implemented by THz phase shifters. Since THz phase shifters are mostly digitally controlled, only quantized angles are available \cite{IndoorTHzComTWC,EEDesignTWC}. We now relax the optimization problem in Section \ref{sec:design_formulation} by adopting the assumption of quantized analog beamformers. Here, we consider the beamsteering codebooks where the codewords have the same form as the antenna array response vectors \cite{AlkhateebFS}.
With the transmit beamsteering codebook at the BS, denoted by $\mathcal{W}$, and the receive beamsteering codebook at the UE, denoted by $\mathcal{V}$, we next design the hybrid beamfoming, including analog beamforming in the RF domain and digital beamforming at the baseband.

\subsection{Analog Beamforming Design}\label{sec:design_codebook_analog}

At the BS, each antenna subarray has one beamsteering direction within the sub-connected architecture. We decompose the channel matrix approximately as $\mathbf{H}[k]=\left[\mathbf{H}_{1}[k],\mathbf{H}_{2}[k],\ldots,\mathbf{H}_{N_\text{RF}}[k]\right]$, where $\mathbf{H}_{n}[k]$ is an $M_rN_r\times M_tN_t$ channel matrix from the $n$th antenna subarray to the UE. Given the target A-AoA, $\theta_{0}^{r}$, and the target E-AoA, $\phi_{0}^{r}$, the corresponding beamsteering vector, denoted by $\mathbf{a}_r(\theta_{0}^{r}, \phi_{0}^{r})$, is adopted as the ideal analog beamformer at the receiver. Similarly, given the target A-AoD, $\theta_{0}^{t}$, and the target E-AoD, $\phi_{0}^{t}$, the corresponding beamsteering vector, denoted by $\mathbf{\hat{a}}_t(\theta_{0}^{t}, \phi_{0}^{t})$, is adopted as the ideal analog beamformer at the transmitter. Hence, the equivalent channel of the subarray at the baseband is expressed as
\begin{align}\label{eq sec4 h_0_uk}
\hat{h}_{n}[k]&=\mathbf{a}^{H}_r(\theta_{0}^{r}, \phi_{0}^{r})\mathbf{H}_{n}[k]\mathbf{\hat{a}}_t(\theta_{0}^{t}, \phi_{0}^{t})\notag\\
&=\alpha^{\textrm{L}}_k G_t G_r \mathcal{A}^{eq}_r(\theta^{r}, \phi^{r})\mathcal{A}^{eq}_t(\theta^{t}, \phi^{t}) P_r(k,\tau)\notag\\
&\quad+ \sum_{i=1}^{N_\textrm{clu}}\sum_{l=1}^{L^{i}_\textrm{ray}} \alpha^{\textrm{NL}}_{k,i,l} G_t G_r \mathcal{A}^{eq}_r(\theta^{r}_{i,l}, \phi^{r}_{i,l}) \mathcal{A}^{eq}_t(\theta^{t}_{i,l}, \phi^{t}_{i,l})P_r(k,\tau_{i,l}),
\end{align}
where
\begin{align} \label{eq sec4 A_eq_r}
&\mathcal{A}^{eq}_r(\theta, \phi) =\frac{1}{\sqrt{M_r N_r}}\sum_{m_r=0}^{M_r-1}\sum_{n_r=0}^{N_r-1} e^{j\frac{2\pi a}{\lambda_c} \left(\Omega_{m_{r},n_{r}}(\theta,\phi)-\Omega_{m_{r},n_{r}}(\theta_{0}^{r},\phi_{0}^{r})\right)}\notag\\
&=\frac{1-e^{j\pi M_r \left(\cos\theta\sin\phi-\cos\theta_{0}^{r}\sin\phi_{0}^{r}\right)}} {\sqrt{M_r N_r}\left(1-e^{j\pi \left(\cos\theta\sin\phi-\cos\theta_{0}^{r}\sin\phi_{0}^{r}\right)}\right)} \times\frac{1-e^{j\pi N_r\left(\sin\theta\sin\phi-\sin\theta_{0}^{r}\sin\phi_{0}^{r}\right)}} {1-e^{j\pi \left(\sin\theta\sin\phi-\sin\theta_{0}^{r}\sin\phi_{0}^{r}\right)}} \notag\\
&\approx\frac{\sin\left[\pi M_r \left(\cos\theta\sin\phi-\cos\theta_{0}^{r}\sin\phi_{0}^{r}\right)\right]}
{\sqrt{M_r N_r}\sin\left[\pi\left(\cos\theta\sin\phi-\cos\theta_{0}^{r}\sin\phi_{0}^{r}\right)\right]} \times \frac{\sin\left[\pi N_r \left(\sin\theta\sin\phi-\sin\theta_{0}^{r}\sin\phi_{0}^{r}\right)\right]}
{\sin\left[\pi\left(\sin\theta\sin\phi-\sin\theta_{0}^{r}\sin\phi_{0}^{r}\right)\right]}
\end{align}
and
\begin{align}\label{eq sec4 A_eq_t}
&\mathcal{A}^{eq}_t(\theta,\phi)=\frac{1}{\sqrt{M_t N_t}}\sum_{m_t=0}^{M_t-1}\sum_{n_t=0}^{N_t-1} e^{j\frac{2\pi a}{\lambda_c} \left(\Omega_{m_{t},n_{t}}(\theta,\phi)-\Omega_{m_{t},n_{t}}(\theta_{0}^{t},\phi_{0}^{t})\right)}\notag\\
&\approx\frac{\sin\left[\pi M_t \left(\cos\theta\sin\phi-\cos\theta_{0}^{t}\sin\phi_{0}^{t}\right)\right]}
{\sqrt{M_t N_t}\sin\left[\pi\left(\cos\theta\sin\phi-\cos\theta_{0}^{t}\sin\phi_{0}^{t}\right)\right]} \times \frac{\sin\left[\pi N_t \left(\sin\theta\sin\phi-\sin\theta_{0}^{t}\sin\phi_{0}^{t}\right)\right]}
{\sin\left[\pi\left(\sin\theta\sin\phi-\sin\theta_{0}^{t}\sin\phi_{0}^{t}\right)\right]}.
\end{align}

We find that when $\theta^{r}$ and $\phi^{r}$ approach the ideal beamforming angles at the receiver, the modulus of $\mathcal{A}^{eq}_r(\theta^r, \phi^r)$ approaches its maximum. When $\theta^{t}$ and $\phi^{t}$ approach the ideal beamforming angles at the transmitter, the modulus of $\mathcal{A}^{eq}_t(\theta^t, \phi^t)$ approaches its maximum. Hence, the optimal beamforming angles can be selected from codebooks, which aims to jointly maximize the sum of normalized equivalent channel modulus over all subcarriers, i.e., $\sum_{k=1}^{K}\frac{\left|\mathbf{a}^{H}_r(\theta^{r}, \phi^{r})\mathbf{H}_{n}[k]\mathbf{\hat{a}}_t(\theta^{t}, \phi^{t})\right|^2}{F(f_k,d)}$. Here, $F(f_k,d)$ is the normalized factor which is equal to the LOS path gain of the $k$th subcarrier channel $\mathbf{H}_{n}[k]$. Also, the effective channel gains are normalized based on center frequencies to ensure the fairness among subcarriers. It is noted that the use of normalized equivalent channel modulus is unique in THz wireless systems and cannot be extended from mmWave or microwave systems. Actually, THz channel gain is approximately deterministic since THz channels are absolutely LOS-dominant and small-scale fading can be ignored. Thus, the BS can estimate the channel gain only based on the transmission frequency and distance. Based on this, we propose the normalized codebook searching algorithm for analog beamforming and present it in \textbf{Algorithm \ref{Alg_CSA}}. For each subarray, we search the given beamsteering codebook to find the optimal transmit beamforming vectors serving all subcarriers. Since the channels of all subcarriers are low-rank and highly correlated, the obtained consistent analog beamformer is expected to work well across all subcarriers. We also need to note that \textbf{Algorithm \ref{Alg_CSA}} is not global optimal. It is impractical to perform the joint searching for each RF chain at both the BS and the UE to find the optimal analog beamforming, since only one RF chain is equipped at the UE. Here, we decouple the joint transmitter-receiver optimization problem and propose this greedy approach. The selected analog beamforming at the UE is optimal on the average sense for all RF chains.

\begin{algorithm}[!t]
\caption{Codebook Searching Algorithm for Analog Beamforming Design}\label{Alg_CSA}
\begin{algorithmic}[1]
\STATE {\bf{Input:}} The receive beamsteering codebook at the UE, $\mathcal{V}$, and the transmit beamsteering codebook at the BS, $\mathcal{W}$.
\STATE Estimate the normalized factor for each subcarrier, $F(f_k,d)=\left(\frac{c}{4\pi{}f_kd}\right)^{2}e^{-k_\textrm{abs}(f_k)d}$.
\STATE Search $\mathcal{V}$ to find the optimal analog beamforming angles at the UE, $\hat{\theta}^{r}$ and $\hat{\phi}^{r}$, such that $\left\{\mathbf{a}_r(\hat{\theta}^{r}, \hat{\phi}^{r})\right\}= \argmax{\sum_{n=1}^{N_\textrm{RF}} \sum_{k=1}^{K}\frac{\left|\mathbf{a}^{H}_r(\theta^{r}, \phi^{r})\mathbf{H}_{n}[k]\right|^2}{F(f_k,d)}}$.
\FOR {$n=1:N_\textrm{RF}$}
\STATE Search $\mathcal{W}$ to find the optimal analog beamforming angles at each RF chain, $\hat{\theta}_{n}^{t}$ and $\hat{\phi}_{n}^{t}$, such that $\left\{\mathbf{\hat{a}}_t(\hat{\theta}_{n}^{t}, \hat{\phi}_{n}^{t})\right\}= \argmax{\sum_{k=1}^{K}\frac{\left|\mathbf{a}^{H}_r(\hat{\theta}^{r}, \hat{\phi}^{r})\mathbf{H}_{n}[k]\mathbf{\hat{a}}_t(\theta^{t}, \phi^{t})\right|^2}{F(f_k,d)}}$.
\ENDFOR
\STATE {\bf{Output:}} $\mathbf{v}=\mathbf{a}_r(\hat{\theta}^{r}, \hat{\phi}^{r})$ and $\mathbf{w}_n= \mathbf{\hat{a}}_t(\hat{\theta}_{n}^{t}, \hat{\phi}_{n}^{t})$,  $n\in\{1,2,\ldots,N_\text{RF}\}$.
\end{algorithmic}
\end{algorithm}

With the obtained analog beamformer, the effective channel at the baseband, $\hat{\mathbf{h}}_{k}$, can be viewed as a MISO channel which is expressed as
\begin{align}\label{eq sec4 h_eq_uk}
&\hat{\mathbf{h}}[k]=\mathbf{v}^{H}\mathbf{H}[k]\mathbf{W}\notag\\
&=\!\mathbf{a}^{H}_r(\hat{\theta}^{r}\!,\!\hat{\phi}^{r})\mathbf{H}[k]\!\begin{bmatrix}
\mathbf{\hat{a}}_t(\hat{\theta}_{1}^{t}, \hat{\phi}_{1}^{t})\!&\!\mathbf{0} \!&\! \cdots \!&\!\mathbf{0}\\
\mathbf{0} \!&\! \mathbf{\hat{a}}_t(\hat{\theta}_{2}^{t}, \hat{\phi}_{2}^{t}) \!&\! \cdots \!&\! \mathbf{0} \\
\vdots \!&\! \vdots \!&\! \vdots \!&\! \vdots \\
\mathbf{0} \!&\! \mathbf{0} \!&\! \cdots \!&\! \mathbf{\hat{a}}_t(\hat{\theta}_{N_\text{RF}}^{t}\!,\! \hat{\phi}_{N_\text{RF}}^{t})
\end{bmatrix}\notag\\
&=\left[\hat{h}_{1}[k],\hat{h}_{2}[k],\ldots,\hat{h}_{N_\text{RF}}[k]\right].
\end{align}
Accordingly, the SINR is rewritten as
\begin{align}\label{eq sec4_gamma_u_k}
\gamma_{k}=\frac{|S_{0}|^2\big|{\hat{\mathbf{h}}[k]}\mathbf{f}_{\textrm{BB}}[k]\big|^2}
{\sum_{\lambda=1,\lambda\neq{}k}^{K}|S_{\lambda-k}|^{2}\big|{\hat{\mathbf{h}}[\lambda]}\mathbf{f}_{\textrm{BB}}[k] \big|^2+\psi}.
\end{align}
Hence, the channel information that needs to be fed back is $\hat{\mathbf{h}}[k]$. This indicates that the exact CSI, $\mathbf{H}[k]$, is no longer needed for digital beamforming design, which significantly reduces the feedback overhead.

\subsection{Digital Beamforming Design}\label{sec:design_codebook_digital}

We now design digital beamforming for IBI elimination. For this design, the effective channels at the baseband, $\{\hat{\mathbf{h}}[k]\}^{K}_{k=1}$, are assumed to be known at the BS through feedback. We adopt the RCI method to eliminate the IBI, which is implemented as \cite{SpatialCCTSP}
\begin{align}\label{eq sec4_f_uk}
\mathbf{f}_{\textrm{BB}}[k]=\left[\left({\mathbf{H}}^{H}_{\text{comb}}[k] {\mathbf{H}}_{\text{comb}}[k] +\beta\mathbf{I}_{N_\text{RF}}\right)^{-1} {\mathbf{H}}^{H}_{\text{comb}}[k]\right]_{:,k},
\end{align}
where $\beta$ is a regularization parameter to be optimized. In \eqref{eq sec4_f_uk}, $\mathbf{H}_{\text{comb}}[k]$ is the combined effective channel given by ${\mathbf{H}}_{\text{comb}}[k]=\left[S_{1-k}\hat{\mathbf{h}}^{T}[1],\ldots, S_{0}\hat{\mathbf{h}}^{T}[k], \ldots, S_{K-k}\hat{\mathbf{h}}^{T}[K]\right]^{T}$.
%\begin{align}\label{eq sec4_H_comb} {\mathbf{H}}_{\text{comb},k}=\left[S_{1-k}\hat{\mathbf{h}}^{T}_{1},\ldots, S_{0}\hat{\mathbf{h}}^{T}_{k}, \ldots, S_{K-k}\hat{\mathbf{h}}^{T}_{K}\right]^{T}.
%\end{align}
The rationale behind using the RCI method is to maximize the SINR optimally. In the following theorem, we show the optimal value of $\beta$ which maximizes the SINR in \eqref{eq sec4_gamma_u_k}.

\begin{theorem}\label{RCI_coefficient_theorem}
$\gamma_{k}$ is maximized when $\beta$ reaches the optimal value, $\beta^{\ast}=\frac{\psi}{\|\mathbf{f}_{\textrm{BB}}[k]\|^{2}_{F}}$.
%\begin{align} \label{eq sec4_beta}
%\beta^{\ast}=\frac{\psi}{\|\mathbf{f}_k\|^{2}_{F}}.
%\end{align}
\begin{IEEEproof}
We first rewrite the SINR as
\begin{align}\label{eq app_bar_gamma_u_k}
\gamma_k&=\frac{|S_{0}|^2 \mathbf{f}^{H}_{\textrm{BB}}[k] \hat{\mathbf{h}}^{H}[k] \hat{\mathbf{h}}[k] \mathbf{f}_{\textrm{BB}}[k]}{\sum_{\lambda=1,\lambda\neq{k}}^{K}|S_{\lambda-k}|^{2}\mathbf{f}^{H}_{\textrm{BB}}[k] \hat{\mathbf{h}}^{H}[\lambda]\hat{\mathbf{h}}[\lambda]\mathbf{f}_{\textrm{BB}}[k]+\psi}\notag\\
&=\frac{|S_{0}|^2 \mathbf{f}^{H}_{\textrm{BB}}[k] \hat{\mathbf{h}}^{H}[k] \hat{\mathbf{h}}[k] \mathbf{f}_{\textrm{BB}}[k]} { \mathbf{f}^{H}_{\textrm{BB}}[k] \left( {\mathbf{H}}^{H}_{\text{comb}}[k]{\mathbf{H}}_{\text{comb}}[k] -|S_{0}|^2 \hat{\mathbf{h}}^{H}[k] \hat{\mathbf{h}}[k] \right)\mathbf{f}_{\textrm{BB}}[k]+\psi}\notag\\
&=\frac{|S_{0}|^2 \mathbf{f}^{H}_{\textrm{BB}}[k] \hat{\mathbf{h}}^{H}[k] \hat{\mathbf{h}}[k] \mathbf{f}_{\textrm{BB}}[k]}
{\mathbf{f}^{H}_{\textrm{BB}}[k]\left({\mathbf{H}}^{H}_{\text{comb}}[k]{\mathbf{H}}_{\text{comb}}[k] +\frac{\psi}{\|\mathbf{f}_{\textrm{BB}}[k]\|^2_{F}}\mathbf{I}_{N_\text{RF}}\right)\mathbf{f}_{\textrm{BB}}[k]-|S_{0}|^2\mathbf{f}^{H}_{\textrm{BB}}[k] \hat{\mathbf{h}}^{H}[k] \hat{\mathbf{h}}[k] \mathbf{f}_{\textrm{BB}}[k]}\notag\\
&=\frac{\omega_k}{1-\omega_k},
\end{align}
where we define $\omega_k$ as
\begin{align}\label{eq app_omega_u_k}
\omega_k=\frac{\mathbf{f}^{H}_{\textrm{BB}}[k] |S_{0}|^2 \hat{\mathbf{h}}^{H}[k] \hat{\mathbf{h}}[k] \mathbf{f}_{\textrm{BB}}[k]}{\mathbf{f}^{H}_{\textrm{BB}}[k]\left({\mathbf{H}}^{H}_{\text{comb}}[k]{\mathbf{H}}_{\text{comb}}[k] +\frac{\psi}{\|\mathbf{f}_{\textrm{BB}}[k]\|^2_{F}}\mathbf{I}_{N_\text{RF}}\right)\mathbf{f}_{\textrm{BB}}[k]}.
\end{align}
Based on \eqref{eq app_omega_u_k}, we find that $0\leq \omega_k<1$ and $\gamma_k$ is a monotonically increasing function of $\omega_k$. Moreover, $\omega_k$ has the form of generalized Rayleigh quotient and the optimal $\mathbf{f}^{H}_{\textrm{BB}}[k]$ that maximizes $\omega_k$ has the same direction as the generalized eigenvector of $\left({\mathbf{H}}^{H}_{\text{comb}}[k]{\mathbf{H}}_{\text{comb}}[k] +\frac{\psi}{\|\mathbf{f}_{\textrm{BB}}[k]\|^2_{F}}\mathbf{I}_{N_\text{RF}}\right)$. Since ${\mathbf{H}}^{H}_{\text{comb}}[k]{\mathbf{H}}_{\text{comb}}[k] +\frac{\psi}{\|\mathbf{f}_{\textrm{BB}}[k]\|^2_{F}}\mathbf{I}_{N_\text{RF}}$ is invertible, the optimal $\mathbf{f}^{H}_{\textrm{BB}}[k]$ that maximizes $\gamma_k$ becomes the dominant eigenvector of $\left({\mathbf{H}}^{H}_{\text{comb}}[k]{\mathbf{H}}_{\text{comb}}[k] +\frac{\psi}{\|\mathbf{f}_{\textrm{BB}}[k]\|^2_{F}}\mathbf{I}_{N_\text{RF}}\right)^{-1}\hat{\mathbf{h}}^{H}[k] \hat{\mathbf{h}}[k]$. Given $\mathbf{f}_{\textrm{BB}}[k]$ in \eqref{eq sec4_f_uk}, the optimal regularization parameter that maximizes SINR is $\beta^{\ast}=\frac{\psi}{\|\mathbf{f}_{\textrm{BB}}[k]\|^{2}_{F}}$, which completes the proof.
\end{IEEEproof}
\end{theorem}

We note that compared to the conventional RCI method, the optimal regularization parameter value in our scheme is normalized by $\|\mathbf{f}_{\textrm{BB}}[k]\|^{2}_{F}$. In our proposed scheme, the analog beamformer is integrated into the combined effective channel, but hybrid beamforming vectors are constrained by the joint power constraint, i.e., $\|\mathbf{W}\mathbf{f}_{\textrm{BB}}[k]\|^{2}_{F}=1$. Hence, the norm of $\mathbf{f}_{\textrm{BB}}[k]$ is not deterministic and needs to be normalized here. In addition, the possible drawback of the aforementioned digital beamforming design may be that the dimension of  $\mathbf{H}_{\text{comb}}[k]$ is too large and thus the computational complexity may not be tolerable. Fortunately, the interference caused by non-adjacent subcarriers can be ignored, since $|S_m|^2$ is close to zero when $m\leq-2$ or $m\geq2$. Hence, we simplify the combined effective channels as
\begin{align}\label{eq sec4_H_eff} \hat{\mathbf{H}}_{\text{comb}}[k]\approx\left[S_{-1}\hat{\mathbf{h}}^{T}[k-1], S_{0}\hat{\mathbf{h}}^{T}[k],S_{1}\hat{\mathbf{h}}^{T}[k+1]\right]^{T}.
\end{align}
Based on \eqref{eq sec4_H_eff}, the original $N_\text{RF}\times K$ channel matrix is simplified as an $N_\text{RF}\times3$ matrix.
When $\mathbf{f}^{H}_{\textrm{BB}}[k]=\left(\hat{\mathbf{H}}^{H}_{\textrm{comb}}[k] \hat{\mathbf{H}}_{\textrm{comb}}[k]+\frac{\psi}{\|\mathbf{f}^{H}_{\textrm{BB}}[k]\|^2_{F}}\mathbf{I}_{N_\text{RF}}\right)^{-1} \hat{\mathbf{h}}^{H}[k]$, the SINR achieves its maximum value, $\gamma_{k}^{\max}$, which is derived as
\begin{align}\label{eq sec4_SINR_max}
\gamma_{k}^{\max}&=\frac{|S_0|^2\hat{\mathbf{h}}[k] \left(\hat{\mathbf{H}}^{H}_{\textrm{comb}}[k] \hat{\mathbf{H}}_{\textrm{comb}}[k]+\psi\mathbf{I}_{N_\text{RF}}\right)^{-1}\hat{\mathbf{h}}^{H}[k] } {1-|S_0|^2\hat{\mathbf{h}}[k]\left(\hat{\mathbf{H}}^{H}_{\textrm{comb}}[k] \hat{\mathbf{H}}_{\textrm{comb}}[k]+\psi\mathbf{I}_{N_\text{RF}}\right)^{-1}\hat{\mathbf{h}}^{H}[k]}\notag\\
&=|S_0|^2\hat{\mathbf{h}}[k] \left(\sum_{\lambda=1,\lambda\neq k}^{K}|S_{\lambda-k}|^2 \hat{\mathbf{h}}^{H}[\lambda] \hat{\mathbf{h}}[\lambda]+ \psi \mathbf{I}_{N_\text{RF}}\right)^{-1} \hat{\mathbf{h}}^{H}[k] \notag\\
&={\rm{tr}}\left\{|S_0|^2\hat{\mathbf{R}}[k] \left(\sum_{\lambda=1,\lambda\neq k}^{K}|S_{\lambda-k}|^2 \hat{\mathbf{R}}[\lambda]+\psi\mathbf{I}_{N_\text{RF}}\right)^{-1}\right\},
\end{align}
where $\hat{\mathbf{R}}[k]$ is defined as $\hat{\mathbf{R}}[k]=\hat{\mathbf{h}}^{H}[k]\hat{\mathbf{h}}[k]$.
With the aforementioned simplification, our proposed hybrid beamforming scheme avoids the processing of large channel matrices and ensures low complexity. This makes our scheme particularly suitable for practical implementation.

\section{Wideband Hybrid Beamforming Scheme with Two Digital Beamformers}\label{sec:design_eig}

In this section, we relax the assumption made in Section \ref{sec:design_codebook} that analog beamformers are selected from beamsteering codebooks.
%Instead, we exploit the EVD of channel covariance matrices in the analog beamforming design.
We note that the hybrid beamforming scheme with beamsteering codebooks, proposed in Section \ref{sec:design_codebook}, is very compatible under the constraints on digital controlled phase shifters. However, an exhaustive search over the codebook is required to find the optimal analog beamforming vectors. This search may be complicated, especially for large codebooks and massive MIMO systems. Hence, in this section we develop a novel wideband hybrid beamforming scheme without using the exhaustive search while achieving high performance for the considered system. Here, we propose to decompose the digital beamforming matrix into two matrices, namely, the compensation matrix and the RCI matrix. The conventional statistical eigen analog beamformer using spatial channel covariance matrices across all subcarriers cannot work well alone in wideband systems due to hardware constraints and the subcarrier channel difference. Thus, the compensation matrix is proposed to assist the analog beamformer and approximate the unconstrained optimal beamformer, while the RCI matrix is used to eliminate the IBI at the baseband. The design of the compensation matrix is challenging due to the low dimensional constraint of digital beamforming matrices in hybrid beamforming systems.

The wideband hybrid beamforming design problem in Section \ref{sec:design_formulation} is a fractional optimization problem, which is in general intractable. The major difficulties in solving the problem are:
\begin{enumerate}
\item The joint optimization of the transmitter and receiver;
\item The coupling between analog and digital beamforming matrices, caused by the power constraint given by $\|\mathbf{W}\mathbf{f}_{\textrm{BB}}[k]\|^{2}_{F}=1$;
\item The non-convex constraints imposed on analog beamformers \cite{gao2018low};
\item The consistent of analog beamformers, $\mathbf{v}$ and $\mathbf{W}$, for all subcarriers.
\end{enumerate}
To conquer the first three difficulties, prior studies in narrow-band hybrid beamforming design \cite{ayach2014spatially,GaoBeamSelection,AALimitedFeedback,Sohrabi2016hybrid,BWangNOMAmmWave} approximated the original problem as a convex one and obtained a near-optimal hybrid beamformer with low complexity. Although these heuristic methods used in such studies were shown to exhibit good performance, they cannot be directly used to design wideband hybrid beamforming in frequency selective THz wireless systems, due to their lack of consideration of the fourth difficulty.
%In \cite{ParkDS}, the fourth difficulty was considered and a near-optimal closed-form hybird beamforming solution was proposed for a mmWave MIMO-OFDM system. However, they assumed fully digital receiver and perfect carrier frequency synchronization, which may be not practical.

Against this background, in this section we tackle all the four difficulties and propose a wideband hybrid beamforming scheme with two digital beamformers, aiming to achieve a comparable performance to the fully digital beamforming design while eliminating the IBI caused by CFO. To this end, we develop a novel design as follows:
\begin{itemize}
\item
First, $\mathbf{W}$ and $\mathbf{v}$ are jointly designed to maximize the desired signal power at the UE using the statistical eigenvalue decomposition (EVD) of spatial channel covariance matrices across all subcarriers, while neglecting the IBI.
\item
Second, we develop $\mathbf{F}_{\textrm{BB},c}[k]$ to compensate for the performance loss of the obtained analog beamformer caused by the constant modulus constraint and the difference among subcarriers.
\item
Third, we develop another digital beamformer, $\mathbf{f}_{\textrm{BB},i}[k]$, to eliminate the IBI, using the obtained hybrid beamformers at the transmitter and the receiver.
\end{itemize}
We note that compared to the mmWave hybrid beamforming scheme in \cite{ParkDS}, we consider a single-RF-chain receiver and imperfect carrier frequency synchronization, which is more practical. Also, a joint analog beamforming design at the transmitter and receiver needs to be considered in our proposed scheme. Most importantly, we develop a hybrid beamforming design strategy with two low dimensional digital beamformers to approximate the optimal beamformer and eliminate interference.

\subsection{Joint Analog Beamforming Design}\label{sec:design_analog_EVD}

Following the aforementioned design strategy, the first step is to jointly optimize $\mathbf{W}$ and $\mathbf{v}$, such that the desired signal power at the UE is maximized. Here, we neglect the IBI.
%Hence, the analog beamforming design problem becomes
%\begin{align}\label{eq sec5_max_R}
%&\left\{\mathbf{v},\mathbf{W}\right\} = \argmax \frac{1}{K}\sum_{k=1}^{K}\left|\mathbf{v}^{H}\mathbf{H}_k \mathbf{W}\right|^2 \notag\\
%&\quad\quad\textrm{s.t.}~~\left|(\mathbf{v})_j\right|=\frac{1}{\sqrt{M_rN_r}},\notag\\
%&\hspace{14mm}\left|(\mathbf{w}_{n})_{i}\right|=\frac{1}{\sqrt{M_tN_t}},
%\end{align}
%where $j\in\{1,2,\ldots,M_rN_r\}$ and $i\in\{1,2,\ldots,M_tN_t\}$.
In narrow-band hybrid beamforming systems, the optimal analog beamformer at the transmitter is composed of eigenvectors of $\mathbf{H}^{H}[k]\mathbf{H}[k]$ and the optimal analog combiner at the receiver is composed of eigenvectors of $\mathbf{H}[k]\mathbf{H}^{H}[k]$ \cite{SpatialCCTSP,yuan2019low}. In frequency selective THz wireless systems, subcarriers have different channels, which results in different analog beamformers. However, the analog beamformer is frequency flat in the considered system, which means that the single analog beamformer needs to serve all subcarriers.
This is an important property of the hybrid beamforming system which distinguishes it from the conventional fully-digital beamforming system \cite{AlkhateebFS}.
Hence, the conventional eigen beamforming method does not work in the considered system since it cannot perform the joint optimization of all subcarriers.

Fortunately, the channel coefficients across subcarriers are highly correlated. We note from \eqref{eq sec2 H_uq} that antenna array response vectors, $\mathbf{a}_r(\theta^{r},\phi^{r})$ and $\mathbf{a}_t(\theta^{t},\phi^{t})$, are the same for all subcarriers.
Moreover, according to \cite{FSohrabiHybridBeamforming}, the eigenvectors of $\mathbf{H}^{H}[k]\mathbf{H}[k]$ are approximately the transmit antenna array response vectors when the number of antennas is sufficiently large.
Thus, the dominant eigenvectors of channel covariance matrices for different subcarriers are approximately the same.
The analog beamforming can be jointly designed with all subcarriers. We propose to incorporate all channel covariance matrices over different subcarriers and design the analog beamforming using a statistical eigen hybrid beamforming scheme, where analog beamforming vectors at the transmitter are designed as the dominant eigenvectors of $\frac{1}{K}\sum_{k=1}^{K}\mathbf{H}^{H}[k]\mathbf{H}[k]$ and analog combining vectors at the receiver are designed as the dominant eigenvectors of $\frac{1}{K}\sum_{k=1}^{K}\mathbf{H}[k]\mathbf{H}^{H}[k]$. Next, we will analyze the asymptotic performance of the proposed scheme compared with the fully digital beamforming scheme and present \textit{Theorem \ref{Eig_theorem}}. For analytical tractability, we focus on the hybrid beamforming at the transmitter and consider the unconstrained analog beamforming.

\begin{theorem}\label{Eig_theorem}
For low-rank wideband correlated THz channels, i.e., the rank of $\frac{1}{K}\sum_{k=1}^{K}\mathbf{H}^{H}[k]\mathbf{H}[k]$ is lower than or equal to the number of RF chains, if the constant modulus constraint is not considered, the proposed statistical eigen hybrid beamforming scheme at the transmitter has approximately the same SINR as that of the fully digital beamforming scheme.
\begin{IEEEproof}
We assume that the rank of $\frac{1}{K}\sum_{k=1}^{K}\mathbf{H}^{H}[k]\mathbf{H}[k]$, $N_\text{H}$, is lower than or equal to the number of RF chains, i.e., $N_\text{H}\leq N_\text{RF}$, and the eigenvectors of $\frac{1}{K}\sum_{k=1}^{K}\mathbf{H}^{H}[k]\mathbf{H}[k]$ associated with its nonzero eigenvalues are $\tilde{\mathbf{U}}$. Thus, $\frac{1}{K}\sum_{k=1}^{K}\mathbf{H}^{H}[k]\mathbf{H}[k]$ can be represented as $\tilde{\mathbf{U}}\tilde{\Lambda}\tilde{\mathbf{U}}^{H}$. We note that $\frac{1}{K}\sum_{k=1}^{K}\mathbf{H}^{H}[k]\mathbf{H}[k]$ is not a full-rank matrix and $\frac{1}{K}\sum_{k=1}^{K}\mathbf{H}^{H}[k]\mathbf{H}[k]=\tilde{\mathbf{U}}\tilde{\Lambda}\tilde{\mathbf{U}}^{H}$ is not a strictly defined EVD. Thus, $\tilde{\Lambda}\in \mathcal{C}^{N_\text{H}\times N_\text{H}}$ is generally not a diagonal matrix.

In the hybrid beamforming design, the analog beamformer without the constant modulus constraint and the sub-connected constraint is given by $\mathbf{W}=[\tilde{\mathbf{U}}~\mathbf{0}]$, which indicates that only $N_\text{H}$ RF chains are effective. Hence, the effective channel covariance matrix is expressed as
\begin{align}\label{eq sec5_Rk_HB}
\hat{\mathbf{R}}^{\text{HB}}&=\frac{1}{K}\sum_{k=1}^{K} \left(\mathbf{H}[k] \mathbf{W}\right)^{H}\mathbf{H}[k] \mathbf{W} \notag\\
&=\mathbf{W}^{H} \left(\frac{1}{K}\sum_{k=1}^{K} \mathbf{H}^{H}[k]\mathbf{H}[k]\right) \mathbf{W} \notag\\
&=\begin{bmatrix}
\tilde{\mathbf{U}}^{H} \\
\mathbf{0}
\end{bmatrix} \tilde{\mathbf{U}}\tilde{\Lambda}\tilde{\mathbf{U}}^{H}\left[\tilde{\mathbf{U}}~\mathbf{0}\right] \notag\\
&=\begin{bmatrix}
\tilde{\Lambda} & \mathbf{0} \\
\mathbf{0} & \mathbf{0}
\end{bmatrix}\in \mathcal{C}^{N_\text{RF}\times N_\text{RF}}.
\end{align}
From \eqref{eq sec4_SINR_max}, the average SINR for all subcarriers with the RCI method is expressed as
\begin{align}\label{eq sec5_SINR_HB}
\gamma^{\text{HB}}&={\rm{tr}}\left\{|S_0|^2 \hat{\mathbf{R}}^{\text{HB}} \left(\sum_{\lambda=0,\lambda\neq k}^{K-1}|S_{\lambda-k}|^2 \hat{\mathbf{R}}^{\text{HB}} +\psi\mathbf{I}_{N_\text{RF}}\right)^{-1}\right\}\notag\\
&={\rm{tr}}\left\{|S_0|^2 \tilde{\Lambda}\left(\sum_{\lambda=0,\lambda\neq k}^{K-1}|S_{\lambda-k}|^2 \tilde{\Lambda}+\psi\mathbf{I}_{N_\text{H}}\right)^{-1}\right\}.
\end{align}

In the fully digital beamforming design, we assume that the eigenvectors of $\mathbf{H}^{H}[k]\mathbf{H}[k]$ associated with its nonzero eigenvalues are $\tilde{\mathbf{U}}[k]$. Let $\mathbf{W}[k]=[\tilde{\mathbf{U}}[k]\quad\mathbf{V}[k]]$ be a unitary matrix, where $\mathbf{V}[k]$ is the null space of $\tilde{\mathbf{U}}[k]$ such that $\tilde{\mathbf{U}}^{H}[k]\mathbf{V}[k]=\mathbf{0}$ and $\mathbf{V}^{H}[k]\mathbf{V}[k]=\mathbf{I}_{N_\text{BS}-N_\text{H}}$. The effective channel covariance matrix is expressed as
\begin{align}\label{eq sec5_Rk_FD}
\hat{\mathbf{R}}^{\text{FD}}&=\frac{1}{K}\sum_{k=1}^{K} \left(\mathbf{H}[k] \mathbf{W}[k]\right)^{H}\mathbf{H}[k] \mathbf{W}[k] \notag\\
&=\frac{1}{K}\sum_{k=1}^{K} \mathbf{W}^{H}[k] \left(\mathbf{H}^{H}[k]\mathbf{H}[k]\right) \mathbf{W}[k] \notag\\
&=\frac{1}{K}\sum_{k=1}^{K} \begin{bmatrix}
\tilde{\mathbf{U}}^{H}[k] \\
\mathbf{V}[k]
\end{bmatrix} \tilde{\mathbf{U}}[k]\tilde{\Lambda}[k]\tilde{\mathbf{U}}^{H}[k] \left[\tilde{\mathbf{U}}[k]~\mathbf{V}[k]\right] \notag\\
&=\frac{1}{K}\sum_{k=1}^{K} \begin{bmatrix}
\tilde{\Lambda}[k] & \mathbf{0} \\
\mathbf{0} & \mathbf{0}
\end{bmatrix}\in \mathcal{C}^{N_\text{BS}\times N_\text{BS}}.
\end{align}
For frequency selective THz channels, we find from \eqref{eq sec2 H_uq} that all subcarriers are highly correlated, i.e., the antenna array response vectors are approximately the same for all subcarriers. %Correspondingly, the eigenvalues of $\mathbf{H}^{H}_{k}\mathbf{H}_{k}$, $\tilde{\Lambda}_k$, are also approximately same for all subcarriers.
Hence, we have $\frac{1}{K}\sum_{k=1}^{K}\tilde{\Lambda}[k]\approx\tilde{\Lambda}$. The average SINR for the fully digital beamforming design is then expressed as
\begin{align}\label{eq sec5_SINR_FD}
\gamma^{\text{FD}}&={\rm{tr}}\left\{|S_0|^2 \hat{\mathbf{R}}^{\text{FD}} \left(\sum_{\lambda=0,\lambda\neq k}^{K-1}|S_{\lambda-k}|^2 \hat{\mathbf{R}}^{\text{FD}} +\psi\mathbf{I}_{N_\text{BS}}\right)^{-1}\right\}\notag\\
&\approx{\rm{tr}}\left\{|S_0|^2 \tilde{\Lambda} \left(\sum_{\lambda=0,\lambda\neq k}^{K-1}|S_{\lambda-k}|^2 \tilde{\Lambda} +\psi\mathbf{I}_{N_\text{H}}\right)^{-1}\right\}.
\end{align}
Since \eqref{eq sec5_SINR_HB} is identical to \eqref{eq sec5_SINR_FD}, the proof is completed.
\end{IEEEproof}
\end{theorem}

We emphasize that the assumption that the rank of $\frac{1}{K}\sum_{k=1}^{K}\mathbf{H}^{H}[k]\mathbf{H}[k]$ is lower than or equal to the number of RF chains is reasonable, since THz channels are low-rank channels and contain a very limited number of NLOS rays. For the sub-connected architecture, however, RF chains are connected to disjoint subsets of antennas, each of which has one unique beamsteering direction \cite{THzSubarrayMag}. Hence, each RF chain has a unique optimal beamforming direction that maximizes the SINR. It follows that we need to design them separately. For the $n$th RF chain, the analog beamformer can be designed as
\begin{align}\label{eq sec5_RF BF}
\mathbf{w}^{\textrm{opt}}_n=\text{eig}_1\left(\frac{1}{K}\sum_{k=1}^{K}\mathbf{H}^{H}_{n}[k] \mathbf{H}_{n}[k]\right),
\end{align}
where $i\in\{1,2,\ldots,M_tN_t\}$. This approach is similar to \cite{liang2014low,yuan2019low} where each RF chain is separated and becomes the basic operational unit. We also note that the practical RF beamformer is not the optimal one, since its amplitude needs to keep constant, constrained by the structure of phase shifters. Hence, we propose that the RF beamformer is designed to have the same angle with the optimal one. The practical RF beamforming vector in the $n$th chain is designed as
\begin{align}\label{eq sec5_w_n}
\left(\mathbf{w}_{n}\right)_{i}=\frac{1}{\sqrt{M_tN_t}}\text{exp}\left(j\measuredangle\left(\left(\mathbf{w}^{\textrm{opt}}_{n}\right)_{i}\right)\right),
\end{align}
where $i\in\{1,2,\ldots,M_tN_t\}$. Similarly, the analog combiner at the receiver is also designed by the statistical eigen hybrid beamforming scheme, which is expressed as
\begin{align}\label{eq sec5_v}
\left(\mathbf{v}\right)_{j}=\frac{1}{\sqrt{M_rN_r}}\text{exp}\left(j\measuredangle\left(\left(\mathbf{v}^{\textrm{opt}}\right)_{j}\right)\right),
\end{align}
where $j\in\{1,2,\ldots,M_rN_r\}$ and $\mathbf{v}^{\textrm{opt}}=\text{eig}_1\left(\frac{1}{K}\sum_{k=1}^{K}\mathbf{H}[k] \mathbf{H}^{H}[k]\right)$. Obviously, the performance of the obtained analog beamformer at the transmitter cannot approach that of the unconstrained fully digital beamformer, due to the concession in \eqref{eq sec5_RF BF} and \eqref{eq sec5_w_n}. This motivates us to design an additional compensation matrix in digital beamforming.

\subsection{Two Digital Beamformers Design}\label{sec:design_transmitter_digital}

In this subsection, we present the digital beamforming design at the transmitter. With the aforementioned design strategy, we propose to decompose the digital beamformer into two low dimensional matrices. The first compensation matrix is used to compensate for the performance loss caused by the difference among subcarriers and hardware constraints. The second RCI matrix is used to eliminate the IBI. Here, we develop the compensation matrix using \emph{Corollary} 2 in \cite{AlkhateebFS}. This baseband beamformer is designed jointly with the analog beamformer at the transmitter to approach the performance of unconstrained fully digital beamforming, which is given by
\begin{align}\label{eq sec5_com BF}
\mathbf{F}_{\textrm{BB},c}[k]=\left(\mathbf{W}^{H}\mathbf{W}\right)^{-\frac{1}{2}}[\bar{\mathbf{V}}[k]]_{:,1:N_{\text{RF}}},
\end{align}
where $\bar{\mathbf{V}}[k]$ is resulting from the singular value decomposition (SVD) of $\mathbf{H}[k]\mathbf{W}\left(\mathbf{W}^{H}\mathbf{W}\right)^{-\frac{1}{2}} = \bar{\mathbf{U}}[k]\bar{\mathbf{\Sigma}}[k]\bar{\mathbf{V}}[k]$. We note that $\mathbf{F}_{\textrm{BB},c}[k]$ is a low dimensional matrix with the dimension of $N_{\text{RF}}\times N_{\text{RF}}$,
but the combination of $\mathbf{W}\mathbf{F}_{\textrm{BB},c}[k]$ is sufficiently close to the optimal digital beamformer, which benefits from the unique sparsity of THz channels.

The RCI baseband beamformer, $\mathbf{f}_{\textrm{BB},i}[k]$, can be obtained using \eqref{eq sec4_f_uk}. The complete digital baseband beamforming vectors are given by $\mathbf{f}_{\text{BB}}[k]=\mathbf{F}_{\textrm{BB},c}[k]\mathbf{f}_{\textrm{BB},i}[k],~~k\in\{1,2,\ldots,K\}$.
%\begin{align}\label{eq sec5_BB BF}
%\mathbf{f}_{\text{BB},k}=\mathbf{F}_{\text{c},k}\mathbf{f}_{i,k},~~k\in\{1,2,\ldots,K\}.
%\end{align}
The overall process for the proposed wideband hybrid beamforming scheme with two digital beamforming matrices is summarized in \textbf{Algorithm \ref{Alg_Eig}}.

\begin{algorithm}[!t]
\caption{Wideband Hybrid Beamforming Scheme with Two Digital Beamformers}\label{Alg_Eig}
\begin{algorithmic}[1]
\STATE {\bf{Input:}} The exact CSI for all subcarriers, $\mathbf{H}[k]$, $k\in\{1,2,\ldots,K\}$.
\FOR {$n=1:N_\text{RF}$}
%\STATE Construct the channel covariance matrices for each subcarrier as $\mathbf{R}_{k,n}=\mathbf{H}^{H}_{k,n}\mathbf{H}_{k,n}$, $k\in\{1,2,\ldots,K\}$.
\STATE Construct the average covariance matrix across all subcarriers as $\bar{\mathbf{R}}_{n}=\frac{1}{K}\sum_{k=1}^{K}\mathbf{H}^{H}_{n}[k]\mathbf{H}_{n}[k]$, $k\in\{1,2,\ldots,K\}$.
\STATE Perform EVD of $\bar{\mathbf{R}}_{n}$ as $\mathbf{U}_n\Lambda_n\mathbf{U}^{H}_n$.
\STATE Construct the RF beamformer on each RF chain as $\mathbf{w}_n=[\mathbf{U}_n]_{:,1}$ and then normalize $\mathbf{w}_n$.
\ENDFOR
\STATE Construct the complete analog beamforming matrix at the transmitter.
\STATE Construct the analog combining matrix at the receiver according to \eqref{eq sec5_v}.
\STATE Construct the compensation baseband beamformer according to \eqref{eq sec5_com BF}.%as $\mathbf{F}_{c,k}=\left(\mathbf{W}^{H}\mathbf{W}\right)^{-\frac{1}{2}}[\bar{\mathbf{V}}_{k}]_{:,1:N_{\text{RF}}}, k\in\{1,2,\ldots,K\}$, where $\bar{\mathbf{V}}_{k}$ is resulting from the SVD of $\mathbf{H}_{k}\mathbf{W}\left(\mathbf{W}^{H}\mathbf{W}\right)^{-\frac{1}{2}} = \bar{\mathbf{U}}_{k}\bar{\mathbf{\Sigma}}_{k}\bar{\mathbf{V}}_{k}$.
\STATE Construct the equivalent channels as $\hat{\mathbf{h}}[k]=\mathbf{v}^{H}\mathbf{H}[k]\mathbf{W}\mathbf{F}_{\textrm{BB},c}[k]$.
\STATE Combine the equivalent channels as $\hat{\mathbf{H}}_{\text{comb}}[k]=\left[S_{-1}\hat{\mathbf{h}}^{T}[k-1], S_{0}\hat{\mathbf{h}}^{T}[k],S_{1}\hat{\mathbf{h}}^{T}[k+1]\right]^{T}$.
\STATE Construct the baseband beamformer for IBI elimination as $\mathbf{f}_{\textrm{BB},i}[k]=\left(\hat{\mathbf{H}}^{H}_{\textrm{comb}}[k] \hat{\mathbf{H}}_{\textrm{comb}}[k]+\frac{\psi}{\|\mathbf{f}_{\textrm{BB}}[k]\|^2_{F}}\mathbf{I}_{N_\text{RF}}\right)^{-1} \hat{\mathbf{h}}^{H}[k]$.
\STATE Construct the complete baseband beamformer as $\mathbf{f}_{\text{BB}}[k]=\mathbf{F}_{\textrm{BB},c}[k]\mathbf{f}_{\textrm{BB},i}[k]$.
\STATE {\bf{Output:}} $\mathbf{v}$, $\mathbf{w}_n$, $n\in\{1,2,\ldots,N_\text{RF}\}$, and $\mathbf{f}_{\text{BB}}[k]$, $k\in\{1,2,\ldots,K\}$.
\end{algorithmic}
\end{algorithm}

Different from \textbf{Algorithm 1}, the design of the analog beamforming in \textbf{Algorithm 2} needs to be followed by updating the compensation digital beamforming, which contains matrix inverse and SVD. However, the optimization in (22) and (23) calculates each vector of the RF beamforming in one stroke, without the need to perform complicated exhaustive searchings. The construction of the RF beamforming only contains matrix multiplication and EVD. We provide the approximated computational complexity of different schemes in Table \ref{Tbl_Compu}, where only the dominant components in calculations are considered. As shown in Table \ref{Tbl_Compu}, the proposed eigen-based hybrid beamforming scheme has lower complexity compared to the proposed codebook searching hybrid scheme in \textbf{Algorithm 1} and the existing hybrid scheme in \cite{EEDesignTWC}, since the exhaustive searching over the codebook is removed. In Table \ref{Tbl_Compu}, $|\mathcal{W}|$ and $|\mathcal{V}|$ denote the codebook size of $\mathcal{W}$ and $\mathcal{V}$, respectively.

\begin{table}[!t]
\centering
\caption{Approximated Computational Complexity of Different Schemes}\label{Tbl_Compu}
\begin{tabular}{|l|c|}
\hline
Hybrid beamforming schemes &Complexity\\
\hline
Codebook-based hybrid scheme &$O(KN^2_\textrm{RF}N_\textrm{BS}N_\textrm{U})+O(KN^3_\textrm{RF})$ \\
Eigen-based hybrid scheme &$O(KN^2_\textrm{RF}|\mathcal{W}|N_\textrm{BS}N_\textrm{U})+O(KN^2_\textrm{RF}|\mathcal{V}|N_\textrm{BS}N_\textrm{U})$ \\
Hybrid scheme in \cite{EEDesignTWC} &$O(KN^2_\textrm{RF}|\mathcal{W}||\mathcal{V}|N_\textrm{BS}N_\textrm{U})$ \\
\hline
\end{tabular}
\end{table}

\section{Robust Hybrid Beamforming Design in the Presence of Imperfect CSI}\label{sec:design_robust}

In Sections \ref{sec:design_codebook} and \ref{sec:design_eig}, we have developed hybrid beamforming schemes under the assumption of perfect CSI. That is, so far we have not considered any channel estimation error in our design. Any non-robust design that does not address the imperfect CSI may degrade the system performance significantly. Against this background, we analyze the impact of imperfect CSI and then design a robust hybrid beamforming scheme using a probabilistic approach to combat estimation errors in this section. The conventional statistical robust approach requires certain statistical assumptions on estimation errors and pays no attention to some extreme cases. The robustness of the adopted probabilistic approach lies in ensuring high quality of service (QoS) performance with high probability even in extreme cases \cite{du2012probabilistic}. In our proposed robust hybrid beamforming scheme, we add a probabilistic constraint to the robust hybrid beamforming design. Instead of maximizing the SINR directly, our scheme aims to maximize average signal power at the receiver and restrain interference power by ensuring a very low probability that the interference power is higher than or equal to the given threshold.

We adopt a frequency-domain compressive channel estimation method proposed for frequency-selective hybrid beamforming systems \cite{J2018frequency}, which very fits our considered system. With this method, the estimated channel matrix can be obtained such that
\begin{align}
\frac{\sum_{k=1}^{K}\|\mathbf{H}[k]-\mathbf{H}_e[k]\|^{2}_{F}}{\sum_{k=1}^{K}\|\mathbf{H}[k]\|^{2}_{F}}\leq \epsilon,
\end{align}
where $\mathbf{H}_e[k]$ is the estimated channel matrix and $\epsilon$ is the normalized mean squared
error (NMSE). In this paper, we focus on the extreme case, i.e., $\frac{\sum_{k=1}^{K}\|\mathbf{H}[k]-\mathbf{H}_e[k]\|^{2}_{F}} {\sum_{k=1}^{K}\|\mathbf{H}[k]\|^{2}_{F}}= \epsilon$. When $K$ is very large, we can obtain $\mathbb{E}(\|\mathbf{H}[k]\|^{2}_{F})\leq (1-\sqrt{\epsilon})\mathbb{E}(\|\mathbf{H}_e[k]\|^{2}_{F})$.

The SINR at the $k$th subcarrier is expressed as
\begin{align}\label{eq sec5 SINR}
\gamma_{k_e} &=\frac{|S_{0}|^2\left|\mathbf{H}[k] \mathbf{W}\mathbf{f}_\textrm{BB}[k]\right|^2}{\sum_{\lambda=1,\lambda\neq{}k}^{K}|S_{\lambda-k}|^{2}
\left|\mathbf{H}[\lambda] \mathbf{W}\mathbf{f}_\textrm{BB}[k]\right|^{2}\!+\!\psi} \notag\\
&=\frac{|S_{0}|^2\left|\mathbf{H}[k] \mathbf{W}\mathbf{f}_\textrm{BB}[k]\right|^2}{\left|\mathbf{\tilde{H}}[k] \mathbf{W}\mathbf{f}_\textrm{BB}[k]\right|^2+\psi}\end{align}
where $\mathbf{\tilde{H}}[k]$ denotes an extended estimated channel matrix that excludes $\mathbf{H}[k]$, i.e., $\mathbf{\tilde{H}}[k]= \left[S_{1-k}\mathbf{H}^{T}[1],\ldots,S_{-1}\mathbf{H}^{T}[k-1], S_{1}\mathbf{H}^{T}[k+1],\ldots,S_{K-k}\mathbf{H}^{T}[K]\right]^{T}$.
With estimation errors on both the numerator and denominator in \eqref{eq sec5 SINR}, it may be difficult to maximize the SINR directly. Hence, our robust scheme takes the probabilistic approach which maximizes average signal power of each subcarrier and ensuring the probability of high interference power very low. Mathematically, the robust hybrid beamforming design problem with the probabilistic constraint is formulated as
\begin{align}\label{eq sec5_max_SINR}
&\hspace{-10mm}\left\{\{\mathbf{W}\},\left\{\mathbf{f}_\textrm{BB}[k]\right\}^{K}_{k=1}\right\} = \argmax \mathbb{E}\left[ \left|\mathbf{H}_e[k] \mathbf{W}\mathbf{f}_\textrm{BB}[k]\right|^2\right] \notag\\
&\quad\quad\quad\textrm{s.t.}~~ Pr \left\{\left|\mathbf{\tilde{H}}[k] \mathbf{W}\mathbf{f}_\textrm{BB}[k]\right|^2\geq T_{k}\right\}\leq p_{k},
\end{align}
where $Pr\{A\}$ denotes the probability of the event $A$, $T_{k}$ denotes a pre-specified interference power threshold, and $p_{k}$ is a given probability.
%The objective function is obtained by taking the expectation of signal power allocated on the $k$th subcarrier with respect to estimation errors, which is presented as
%\begin{align}\label{eq sec6_Objec}
%&\mathbb{E}\left[ \left|S_0(\mathbf{H}_{k_p}+\mathbf{E}_{k}) \mathbf{W}\mathbf{f}_k\right|^2\right]\notag\\
%=&|S_0|^2\mathbb{E}\left[(\mathbf{W}\mathbf{f}_k)^{H} (\mathbf{H}_{k_p}+\mathbf{E}_{k})^{H}(\mathbf{H}_{k_p}+\mathbf{E}_{k})\mathbf{W}\mathbf{f}_k\right] \notag\\
%=&|S_0|^2\mathbb{E}\left[{\rm{tr}}\left\{\left(\mathbf{H}^{H}_{k}\mathbf{H}_{k}+\mathbf{E}^{H}_{k}\mathbf{E}_{k}\right) \mathbf{M}_k \right\}\right]\notag\\
%=&|S_0|^2{\rm{tr}}\left\{\left(\mathbf{H}^{H}_{k_p}\mathbf{H}_{k_p}+\sigma^2_{E}\mathbf{I}_{N_\text{BS}}\right)\mathbf{M}_k\right\},
%\end{align}
%where $\mathbf{M}_k\triangleq(\mathbf{W}\mathbf{f}_k)(\mathbf{W}\mathbf{f}_k)^{H}$ is a symmetric positive semi-definite matrix.
Here, we introduce the constraint in to guarantee that the probability for the interference power of the $k$th subcarrier to exceed a pre-specified threshold $T_{k}$ is less than $p_{k}$. In order to design hybrid beamforming, the major challenge now is to convert the probabilistic constraint into a deterministic form. We define a non-negative random variable $Z_k=(\mathbf{W}\mathbf{f}_\textrm{BB}[k])^{H}\mathbf{\tilde{H}}^{H}[k] \mathbf{\tilde{H}}[k]\mathbf{W}\mathbf{f}_\textrm{BB}[k]$. Using Markov's inequality, the left hand side of the constraint satisfies
\begin{align}\label{eq sec6_Z_k}
Pr \left\{Z_k\geq T_{k}\right\}\leq\frac{\mathbb{E}\left[Z_k\right]}{T_{k}}.
\end{align}
Taking expectation with respect to different channel realizations, we obtain $\mathbb{E}\left[Z_k\right]$ as
\begin{align}\label{eq sec6_expect_Z_k}
\mathbb{E}\left[Z_k\right] &= \mathbb{E}\left[{\rm{tr}}\left\{\mathbf{\tilde{H}}^{H}[k] \mathbf{\tilde{H}}[k]\mathbf{W}\mathbf{f}_\textrm{BB}[k]\left(\mathbf{W}\mathbf{f}_\textrm{BB}[k]\right)^{H} \right\}\right] \notag\\
&\leq{\rm{tr}}\left\{(1-\sqrt{\epsilon})\mathbf{\tilde{H}}^{H}_e[k] \mathbf{\tilde{H}}_e[k] \mathbf{W}\mathbf{f}_\textrm{BB}[k]\left(\mathbf{W}\mathbf{f}_\textrm{BB}[k]\right)^{H}\right\}.
\end{align}
With \eqref{eq sec6_Z_k} and \eqref{eq sec6_expect_Z_k}, it can be easily verified that the probabilistic constraint is satisfied if ${\rm{tr}}\left\{(1-\sqrt{\epsilon})\mathbf{\tilde{H}}^{H}_e[k] \mathbf{\tilde{H}}_e[k] \mathbf{W}\mathbf{f}_\textrm{BB}[k]\left(\mathbf{W}\mathbf{f}_\textrm{BB}[k]\right)^{H}\right\}\leq p_k T_{k}$.
Hence, the probabilistic problem in \eqref{eq sec5_max_SINR} can be transformed into the following form by reformulating the objective function and the probabilistic constraint, which is presented by
\begin{align}\label{eq sec5_max_SINR2}
&\hspace{-10mm}\left\{\mathbf{M}[k]\right\} = \argmax {\rm{tr}}\left\{\mathbf{H}^{H}_e[k] \mathbf{H}_e[k] \mathbf{M}[k]\right\} \notag\\
&\quad\quad\textrm{s.t.}~~ {\rm{tr}}\left\{\mathbf{\tilde{H}}^{H}_e[k] \mathbf{\tilde{H}}_e[k] \mathbf{M}[k]\right\}\leq \frac{p_k T_{k}}{1-\sqrt{\epsilon}},
\end{align}
where $\mathbf{M}[k]=\mathbf{W}\mathbf{f}_\textrm{BB}[k]\left(\mathbf{W}\mathbf{f}_\textrm{BB}[k]\right)^{H}$ is a matrix to be optimized. We emphasize that only the expectation of estimated channel covariance matrices, i.e., $\mathbb{E}\left[\mathbf{H}_{e}[k]^{H}\mathbf{H}_{e}[k]\right]$, is required at the BS. This indicates that the exact estimation error variances for all subcarriers are no longer needed for our robust scheme, which is more practical than the conventional statistical scheme. The optimal solution can be obtained by a standard mathematical programming $\textit{CVX}$ in \cite{grantCVX}. We note that the objective function is a convex expression since ${\rm{tr}}\left\{\mathbf{H}^{H}[k] \mathbf{H}[k] \mathbf{M}[k]\right\}$ is a complex value. However, maximizing a convex expression in the objective function cannot be solved in the $\textit{CVX}$. Here, we propose a greedy separate optimization method where the amplitude and the angle of complex entries in $\mathbf{M}[k]$ are optimized separately. First, we propose to solve the problem in \eqref{eq sec5_max_SINR2} only considering the amplitude information and the obtained optimal solution with $\textit{CVX}$, $\mathbf{M}^{opt}[k]$, is a real matrix. Then, the angle information of entries in $\mathbf{M}^{opt}[k]$ is optimized by applying randomization technique. The idea of the randomization technique is to generate a set of candidate complex vectors using $\mathbf{M}^{opt}[k]$ and then choose the best solution from these candidate complex vectors. More specifically, we calculate the EVD of $\mathbf{M}^{opt}[k]=\hat{\mathbf{U}}[k]\hat{\Lambda}[k]\hat{\mathbf{U}}[k]^{H}$ and obtain the candidate complex vectors by $\mathbf{f}^{cand}_{i}[k]=\hat{\mathbf{U}}[k]\hat{\Lambda}[k]^{1/2}\mathbf{v}_{i}$, where $\mathbf{v}_{i}$ is a zero-mean complex Gaussian vector whose covariance matrix is $\mathbf{I}$.
It is necessary to check whether the constraint are violated by the candidate vector. The optimal hybrid beamforming vector for each subcarrier, $\mathbf{f}^{opt}_\textrm{hybrid}[k]$, is selected among those candidate complex vectors, which has the smallest norm. With $\mathbf{f}^{opt}_\textrm{hybrid}[k]$, we can recover the optimal analog beamforming matrix and digital beamforming vectors using the method in Section \ref{sec:design_eig}.

\section{Numerical Results and Discussions}\label{sec:results}

In this section, we evaluate the spectral efficiencies of our proposed hybrid beamforming schemes. To demonstrate the benefits of our proposed schemes, we consider two benchmark schemes: i) The unconstrained fully digital beamforming scheme with no IBI and ii) The existing hybrid beamforming scheme proposed in \cite{EEDesignTWC}.
In THz communications, highly directional antennas and high antenna gains are inevitable due to the high path loss. Throughout this section, all simulation parameters are listed in Table II, where the molecular absorption coefficient is calculated by the HITRAN database. In each channel realization, the azimuth AoAs and AoDs of the LOS path are assumed to be uniformly distributed in $[0, 2\pi]$ and the elevation AoAs and AoDs of the LOS path are uniformly distributed in $[-\pi/2, \pi/2]$.

\begin{table}[!t]
\centering
\caption{Simulation Parameters}\label{Tbl_Simu}
\begin{tabular}{|l|c|}
\hline
Parameters & Values \\
\hline
Operating frequency & $300$ GHz to $450$ GHz\\
Antenna beam-width  & $20^{\circ}$\\
Transmit antenna gain & $20$ dBi\\
Receive antenna gain  & $20$ dBi\\
Bandwidth  & $1$ GHz\\
Noise power  & $-75~\textrm{dBm}$\\
Fresnel reflection coefficient & 0.15\\
Number of subcarriers & 128\\
Cyclic prefix length & 16\\
Number of NLOS paths & 3\\
CFO ratio & 0.3\\
codebook size & 3 bits\\
Sampling time & $7.8\times10^{-12}$s\\
\hline
\end{tabular}
\end{table}

\begin{figure}[!t]
    \begin{center}
    \includegraphics[height=3.2in,width=0.7\columnwidth]{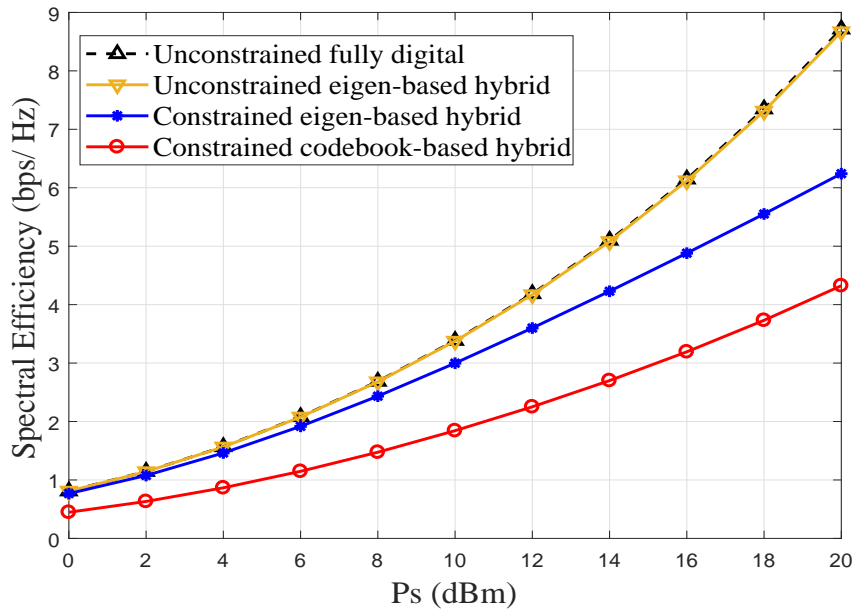}
    \caption{Spectral efficiencies of proposed hybrid beamforming schemes versus $P_s$ under the assumption of no IBI. $d=5~\textrm{m}$. $N_\text{RF}=4$, $M_t\times N_t=8\times8=64$, and $M_r\times N_r=8\times8=64$.}\label{Fig:SEvsPsnoi}
    \end{center}\vspace{-4mm}
\end{figure}

Fig. \ref{Fig:SEvsPsnoi} plots the spectral efficiencies of our proposed hybrid beamforming schemes versus $P_{s}$. In this figure, we assume an ideal receiver (fully digital receiver) and no IBI.
We first observe that the performance of the unconstrained hybrid beamforming scheme is very close to that of the unconstrained fully digital beamforming scheme, which demonstrates the accuracy of \textit{Theorem \ref{Eig_theorem}}. Second, we observe that our proposed hybrid beamforming schemes can achieve half to 2/3 of the spectral efficiency of the unconstrained fully digital beamforming scheme. Considering that our proposed schemes have lower hardware complexity than the unconstrained fully digital beamforming scheme, this observation demonstrates the significant benefit of our proposed schemes, especially for the sake of practical implementation. Third, our proposed eigen-based hybrid beamforming scheme significantly outperforms our proposed codebook-based hybrid beamforming scheme. This performance loss associated with the codebook-based hybrid beamforming scheme comes from the quantification on phase shifters. Although the codebook-based hybrid beamforming scheme is not optimal for spectral efficiency maximization, it is an attractive alternative since phase shifters are mostly digitally controlled in the THz band.

\begin{figure}[!t]
    \begin{center}
    \includegraphics[height=3.2in,width=0.7\columnwidth]{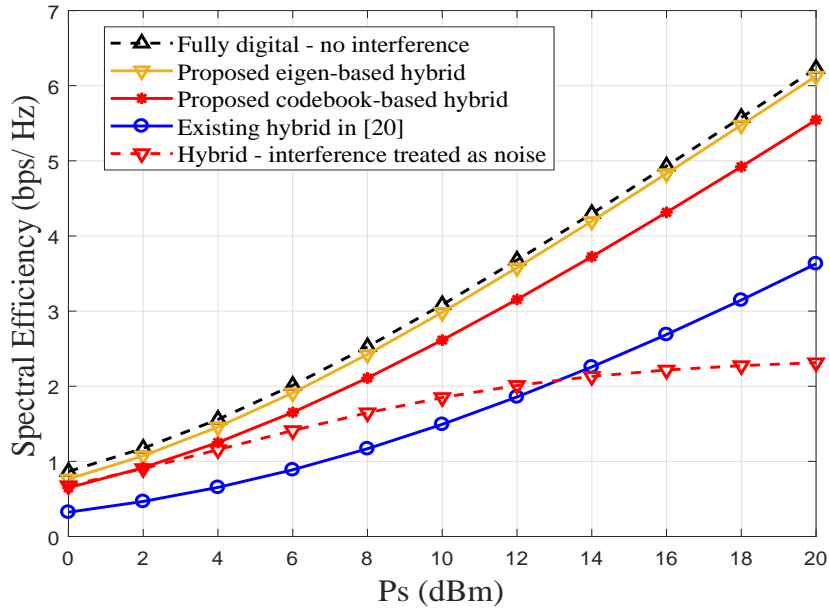}
    \caption{Spectral efficiencies of proposed hybrid beamforming schemes versus $P_s$ with the IBI. $d=5~\textrm{m}$. $N_\text{RF}=4$, $M_t\times N_t=8\times8=64$, and $M_r\times N_r=8\times8=64$.}\label{Fig:SEvsPsi}
    \end{center}\vspace{-4mm}
\end{figure}

Fig. \ref{Fig:SEvsPsi} plots the spectral efficiencies of three hybrid beamforming schemes with IBI versus $P_{s}$. For the fairness of comparison, the receivers in all schemes are set as the same analog combiner, considering that only one RF chain is equipped at the UE. We also plot the spectral efficiency of the proposed eigen-based hybrid beamforming scheme where the IBI is treated as noise, i.e., no interference elimination, for comparison. We first observe from Fig. \ref{Fig:SEvsPsi} that our proposed hybrid beamforming schemes with IBI elimination significantly outperform the existing hybrid beamforming scheme proposed in \cite{EEDesignTWC}. This observation reveals the advantage of our proposed hybrid beamforming schemes, especially over frequency selective fading. We also observe that the spectral efficiency of the proposed eigen-based hybrid scheme with the RCI method is much higher than that of the scheme without IBI elimination, especially when the transmit power is high. This indicates that the IBI has a significant impact on the system performance and the RCI method is effective to eliminate the IBI.

\begin{figure}[!t]
    \begin{center}
    \includegraphics[height=3.2in,width=0.7\columnwidth]{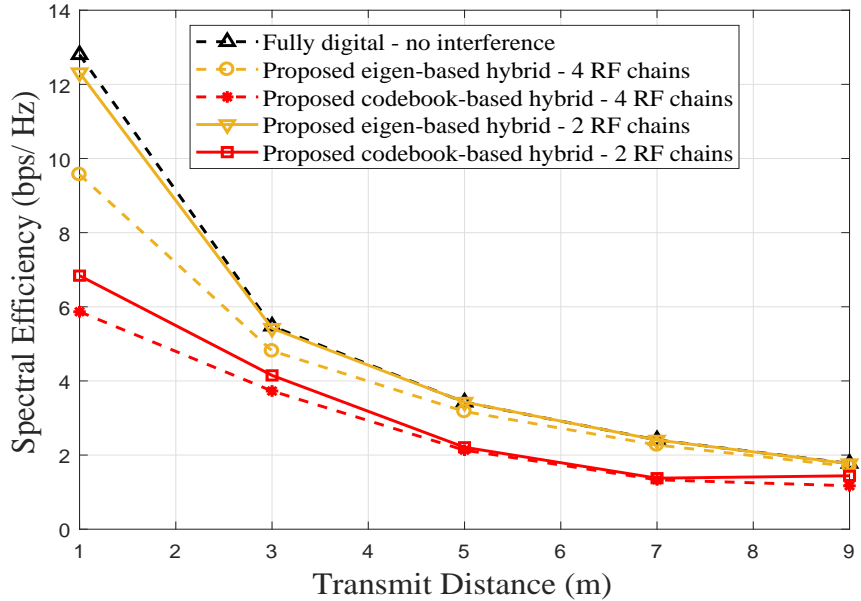}
    \caption{Spectral efficiencies of three schemes versus $d$ with different $N_\text{RF}$. $P_s=10~\textrm{dBm}$. $N_\text{BS}=256$ and $M_r\times N_r=8\times8=64$. }\label{Fig:SEvsD}
    \end{center}\vspace{-4mm}
\end{figure}

Fig. \ref{Fig:SEvsD} plots the spectral efficiencies of our proposed hybrid beamforming schemes versus $d$. Fig. \ref{Fig:SEvsD} shows that the spectral efficiencies decrease rapidly as the BS-UE distance increases, which is resulting from the severe path loss in THz channels. Although a massive number of antennas are equipped at the BS, the maximum effective transmission distance is generally limited to $10~\textrm{m}$. We also plot the spectral efficiencies of our proposed hybrid beamforming schemes with different $N_\text{RF}$. While $N_\text{BS}$ is fixed, the number of RF chains is set as $N_\text{RF}=2$ or $N_\text{RF}=4$. We observe that the spectral efficiencies of our proposed hybrid beamforming schemes with $4$ RF chains are higher than those with $2$ RF chains. Specifically, the performance of our proposed statistical eigen hybrid beamforming scheme is very close to that of the unconstrained fully digital beamforming scheme when $N_\text{RF}=4$. However, when $N_\text{RF}=2$, the performance of our proposed hybrid beamforming schemes is much lower than that of the unconstrained fully digital beamforming scheme. Considering that the number of NLOS channel paths is $3$, this observation demonstrates that $N_\text{RF}$ need to be higher than or equal to the rank of THz channels.

\begin{figure}[!t]
    \begin{center}
    \includegraphics[height=3.2in,width=0.7\columnwidth]{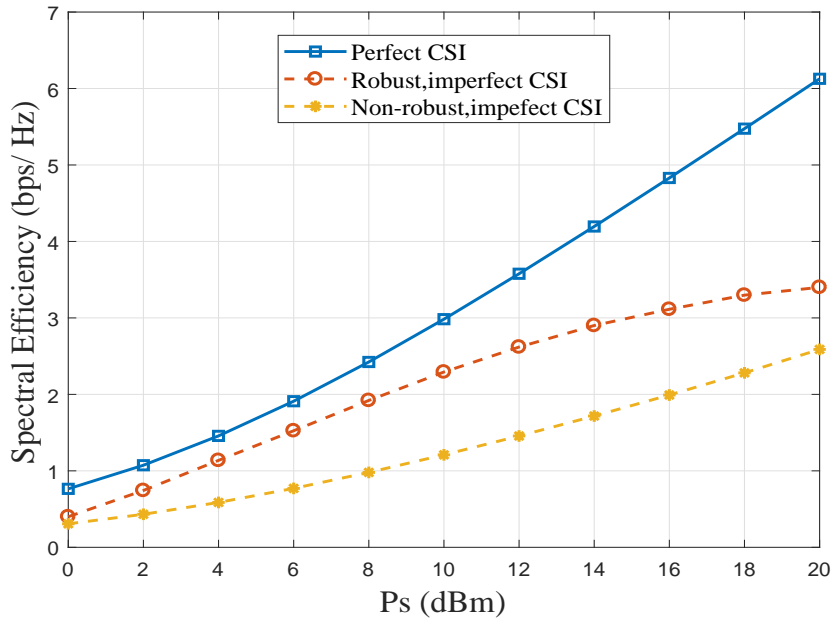}
    \caption{Spectral efficiencies of the proposed robust hybrid beamforming scheme versus the total transmit power in the presence of imperfect CSI. The transmit distance between the BS and the UE is $5~\textrm{m}$. $N_\text{RF}=4$, $M_t\times N_t=8\times8=64$, and $M_r\times N_r=8\times8=64$.}\label{Fig:SE_imperfect_CSI}
    \end{center}\vspace{-4mm}
\end{figure}

Fig. \ref{Fig:SE_imperfect_CSI} plots the spectral efficiencies of the proposed wideband hybrid beamforming scheme with two digital beamforming matrices versus $P_{s}$. In this figure, we compare the performance in three cases. In the first case, the perfect CSI is assumed to be known at the BS. In the second and third cases, only imperfect CSI is known at the BS where the estimation NMSE is $\epsilon=0.002$. The proposed robust hybrid beamforming scheme is adopted to mitigate the channel estimation error in the second case, where the interference power threshold in the probabilistic constraint is set as $T_{k} = 1\times10^{-8}$ and the probability threshold is set as $p_{k}=5\%$. The non-robust hybrid beamforming scheme is used in the third case. We observe that the proposed robust hybrid beamforming scheme provides a clear performance advantage over the non-robust one. When $P_{s}\leq 14~\textrm{dBm}$, the performance gap between the robust scheme and the non-robust scheme increases with $P_{s}$. This is because that the estimation NMSE is assumed to be independent of the transmit power. When $P_s$ is low, the interference can be effectively controlled by our proposed probabilistic approach and the Gaussian noise plays a major role for the performance loss. However, when $P_{s}$ becomes very high, the performance gap between the robust scheme and the non-robust scheme decreases slightly. This is because when $P_s$ is very high, the detrimental impact of the IBI becomes extreme severe. The probabilistic constraint might be too strict and the probabilistic approach cannot eliminate the interference, which is limited by the performance of the mathematical programming. Thus, the IBI plays an important role for the performance loss in the extreme high transmit power regime.

\section{Conclusion}\label{sec:con}

In this work, we proposed new wideband hybrid beamforming schemes for the THz multi-carrier system over frequency selective fading. Importantly, we considered the CFO and unique THz channel  characteristics in the system modeling and design. We first built a 3D THz channel model to characterize the frequency selective propagation in an indoor environment. Based on this model, we proposed a two-stage wideband hybrid beamforming scheme using a beamsteering codebook searching algorithm for analog beamforming and a revised RCI method for digital beamforming. To avoid the exhaustive search over codebooks, we proposed a novel wideband hybrid beamforming scheme with two digital beamformers using the EVD of channel covariance matrices. Furthermore, we analyzed the impact of channel estimation errors and then designed a robust hybrid beamforming scheme to combat the imperfect CSI. Using numerical results, we showed that our proposed schemes exhibit a comparable spectral efficiency to the fully digital beamforming. Also, our proposed schemes achieve a significantly higher spectral efficiency than the existing hybrid beamforming scheme. Thus, our design reveals the profound practicality to employ frequency selective hybrid beamforming in THz wireless communications.


\begin{thebibliography}{10}

\bibitem{yuan2018hybrid}
H. Yuan, N. Yang, K. Yang, C. Han, and J. An,
\newblock ``Hybrid beamforming for MIMO-OFDM terahertz wireless systems over frequency selective channels,''
\newblock in {\em Proc. IEEE GLOBECOM}, Abu Dhabi, United Arab Emirates, Dec. 2018, pp. 1--6.

\bibitem{Federici2010review}
J. Federici and L. Moeller,
\newblock ``Review of terahertz and subterahertz wireless communications,''
\newblock {\em J. Appl. Phys.}, vol. 107, pp. 101--111, Mar. 2010.

\bibitem{akyildiz2014terahertz}
I. F. Akyildiz, J. M. Jornet, and C. Han,
\newblock ``Terahertz band: Next frontier for wireless communications,''
\newblock {\em Phys. Commun.}, vol. 12, no. 2, pp. 16–-32, Sep. 2014.

\bibitem{akyildiz2010electromagnetic}
I. F. Akyildiz and J. M. Jornet,
\newblock ``Electromagnetic wireless nanosensor networks,''
\newblock {\em Nano Commun. Networks J.}, vol. 1, no. 1, pp. 3–-19, Mar. 2010.

\bibitem{akyildiz2014teranets}
I. F. Akyildiz, J. M. Jornet, and C. Han,
\newblock ``TeraNets: Ultra-broadband communication networks in the terahertz band,''
\newblock {\em IEEE Wireless Commun. Mag.}, vol. 21, no. 4, pp. 130–-135, Aug. 2014.

\bibitem{jornet2012phlame}
J. M. Jornet, J. C. Pujol, and J. S. Pareta,
\newblock ``Phlame: A physical layer aware mac protocol for electromagnetic nanonetworks in the terahertz band,''
\newblock {\em Nano Commun. Networks J.}, vol. 3, no. 1, pp. 74–-81, Mar. 2012.

\bibitem{an2020energy}
J. An, Y. Zhang, X. Gao, and K. Yang,
\newblock ``Energy-efficient base station association and beamforming for multi-cell multiuser systems,''
\newblock {\em IEEE Trans. Wireless Commun.}, early access.

\bibitem{gao2020contract}
X. Gao, D. Niyato, P. Wang, K. Yang, and J. An,
\newblock ``Contract design for time resource assignment and pricing in backscatter-assisted RF-powered networks,''
\newblock {\em IEEE Wireless Commun. Lett.}, vol. 9, no. 1, pp. 42--46, Jan. 2020.

%
%\bibitem{gao2019auction}
%X. Gao, P. Wang, D. Niyato, K. Yang, and J. An,
%\newblock ``Auction-Based Time Scheduling for Backscatter-Aided RF-Powered Cognitive Radio Networks,''
%\newblock {\em IEEE Trans. Wireless Commun.}, vol. 18, no. 3, pp. 1684–-1697, Mar. 2019.
%
%\bibitem{RappaportmmWave5G}
%T. S. Rappaport {\em et al.},
%\newblock ``Millimeter wave mobile communications for 5G cellular: It will work!''
%\newblock {\em IEEE Access}, vol. 1, pp. 335--349, May 2013.

\bibitem{Heathoverview}
R. W. Heath, Jr., N. Gonz\'{a}lez-Prelcic, S. Rangan, W. Roh, and A. M. Sayeed,
\newblock ``An overview of signal processing techniques for millimeter wave MIMO systems,''
\newblock {\em IEEE J. Sel. Topics Signal Process.}, vol. 10, no. 3, pp. 436--453, Apr. 2016.

\bibitem{THzSubarrayMag}
C. Lin and G. Y. Li,
\newblock ``Terahertz communications: An array-of-subarrays solution,''
\newblock {\em IEEE Commun. Mag.}, vol. 54, no. 12, pp. 124--131, Dec. 2016.

\bibitem{rappaport2011state}
T. Rappaport, J. Murdock, and F. Gutierrez,
\newblock ``State of the art in 60-GHz integrated circuits and systems for wireless communications,''
\newblock {\em Proc. IEEE}, vol. 99, no. 8, pp. 1390–-1436, Aug. 2011.

\bibitem{ju2011graphene}
L. Ju, B. Geng, J. Horng, C. Girit, M. martin, Z. Hao, H. Bechtel, X. Liang, A. Zettl, Y.R. Shen, and F. Wang,
\newblock ``Graphene plasmonics for tunable terahertz metamaterials,''
\newblock {\em Nature Nanotechnol.}, vol. 6, no. 10, pp. 630–-634, Oct. 2011.

\bibitem{liu2010broadband}
L. Liu, J. Hesler, H. Xu, A. Lichtenberger, and R. Weikle,
\newblock ``A broadband quasi-optical terahertz detector utilizing a zero bias Schottky diode,''
\newblock {\em IEEE Microw. Wirel. Compon. Lett.}, vol. 20, no. 9, pp. 504-–506, Sep. 2010.

\bibitem{han2018ultra}
C. Han, J. M. Jornet, and I. F. Akyildiz,
\newblock ``Ultra-massive MIMO channel modeling for graphene-enabled terahertz-band communications,''
\newblock in {\em 2018 Proc. IEEE Veh. Technol. Conf. (VTC)}, Jun. 2018, pp. 1--5.

\bibitem{Sarieddeen2019tera}
H. Sarieddeen, M.-S. Alouini, and T. Y. Al-Naffouri,
\newblock ``Terahertz-band ultra-massive spatial modulation MIMO,''
\newblock {\em IEEE J. Select. Areas Commun.}, vol. 37, no. 9, pp. 2040--2052, Sept. 2019.

\bibitem{jornet2013graphene}
J. M. Jornet and I. F. Akyildiz,
\newblock ``Graphene-based plasmonic nano-antenna for terahertz band communication in nano-networks,''
\newblock {\em J. Select. Areas Commun.}, vol. 12, no. 12, pp. 685--694, Dec. 2013.

%\bibitem{jornet2014graphene}
%J.  M.  Jornet and I. F. Akyildiz,
%\newblock ``Graphene-based plasmonic nano-transceiver for terahertz band communication,''
%\newblock {\em in Proc. 8th EuCAP}, The Hague, Netherlands, Apr. 2014, pp. 492--496.

\bibitem{Sengupta2018tera}
K. Sengupta, T. Nagatsuma, and D. Mittleman,
\newblock ``Terahertz integrated electronic and hybrid electronic-photonic systems,'' \newblock {\em Nature Electron.}, vol. 1, no. 12, pp. 622--635, 2018.


\bibitem{zhang2019mixed}
J. Zhang, L. Dai, Z. He, B. Ai, and O. A. Dobre,
\newblock ``Mixed-ADC/DAC multipair massive MIMO relaying systems: Performance analysis and power optimization,''
\newblock {\em IEEE Trans. Commun.}, vol. 67, no. 1, pp. 140--153, Jan. 2019.


\bibitem{IndoorTHzComTWC}
C. Lin and G. Y. Li,
\newblock ``Indoor terahertz communications: How many antenna arrays are needed?''
\newblock {\em IEEE Trans. Wireless Commun.}, vol. 14, no. 6, pp. 3097--3107, June 2015.

%\bibitem{AdaptiveBFTCOM}
%C. Lin and G. Y. Li,
%\newblock ``Adaptive beamforming with resource allocation for distance-aware multi-user indoor terahertz communications,''
%\newblock {\em IEEE Trans. Commun.}, vol. 63, no. 8, pp. 2985--2995, Aug. 2015.

\bibitem{EEDesignTWC}
C. Lin and G. Y. Li,
\newblock ``Energy-efficient design of indoor mmWave and sub-THz systems with antenna arrays,''
\newblock {\em IEEE Trans. Wireless Commun.}, vol. 15, no. 7, pp. 4660--4672, July 2016.

\bibitem{WidebandWaveform}
C. Han, A. O. Bicen, and I. F. Akyildiz,
\newblock ``Multi-wideband waveform design for distance-adaptive wireless communications in the terahertz band,''
\newblock {\em IEEE Trans. Signal Process.}, vol. 64, no. 4, pp. 910--922, Feb. 2016.

\bibitem{wang2018spatial}
B. Wang, F. Gao, S. Jin, H. Lin and G. Y. Li,
\newblock ``Spatial- and frequency-wideband effects in millimeter-wave massive MIMO systems''
\newblock {\em IEEE Trans. Signal Process.}, vol. 66, no. 13, pp. 3393--3406, July, 2018.

\bibitem{wang2019beam}
B. Wang, M. Jian, F. Gao, G. Y. Li and H. Lin,
\newblock ``Beam squint and channel estimation for wideband mmWave massive MIMO-OFDM Systems''
\newblock {\em IEEE Trans. Signal Process.}, vol. 67, no. 23, pp. 5893--5908, Dec. 2019.

\bibitem{AlkhateebFS}
A. Alkhateeb and R. W. Heath,
\newblock ``Frequency selective hybrid precoding for limited feedback millimeter wave systems,''
\newblock {\em IEEE Trans. Commun.}, vol. 64, no. 5, pp. 1801--1818, May 2016.

\bibitem{ParkDS}
S. Park, A. Alkhateeb, and R. W. Heath,
\newblock ``Dynamic subarrays for hybrid precoding in wideband mmWave MIMO systems,''
\newblock {\em IEEE Trans. Wireless Commun.}, vol. 16, no. 5, pp. 2907--2920, May 2017.

\bibitem{FSohrabiHybridBeamforming}
F. Sohrabi and W. Yu,
\newblock ``Hybrid analog and digital beamforming for mmWave OFDM large-scale antenna arrays,''
\newblock {\em IEEE J. Select. Areas Commun.}, vol. 35, no. 7, July 2017.

\bibitem{you2017BDMA}
L. You, X. Gao, G. Y. Li, X.-G. Xia, and N. Ma,
\newblock ``BDMA for millimeter-wave/terahertz massive MIMO transmission with per-beam synchronization,''
\newblock {\em IEEE J. Sel. Areas Commun.}, vol. 35, no. 7, pp. 1550--1563, Jul. 2017.

\bibitem{DABA}
C. Han and I. F. Akyildiz,
\newblock ``Distance-aware bandwidth-adaptive resource allocation in the terahertz band,''
\newblock {\em IEEE Trans. Terahertz Sci. Technol.}, vol. 6, no. 4, pp. 541--553, July 2016.

\bibitem{chen2013tera}
P.-Y. Chen, C. Argyropoulos, and A. Alu,
\newblock ``Terahertz antenna phase shifters using integrally-gated graphene transmission-lines,''
\newblock {\em IEEE Trans. Antennas Propag.}, vol. 61, no. 4, pp. 1528--1537, Apr. 2013.

\bibitem{wu2013graphene}
Y. Wu {\em et al.},
\newblock ``Graphene/liquid crystal based Terahertz phase shifters,''
\newblock {\em Opt. Exp.}, vol. 21, no. 18, pp. 21395--21402, Sep. 2013.

%\bibitem{alice2019ultra}
%F. Alice, H. Sarieddeen, H. Dahrouj, T. Y. Al-Naffouri, and M.-S. Alouini,
%\newblock ``Ultra-massive MIMO systems at terahertz bands: Prospects and challenges,''
%\newblock {\em arXiv preprint}, arXiv:1902.11090 (2019).

%\bibitem{Han2014distance}
%C. Han and I. F. Akyildiz,
%\newblock ``Distance-aware multi-carrier (DAMC) modulation in terahertz band communication,''
%\newblock in {\em Proc. IEEE ICC}, Sydney, Australia, June 2014, pp. 5461--5467.

\bibitem{ChannelModelTWC}
J. M. Jornet and I. F. Akyildiz,
\newblock ``Channel modeling and capacity analysis for electromagnetic wireless nanonetworks in the terahertz band,''
\newblock {\em IEEE Trans. Wireless Commun.}, vol. 10, no. 10, pp. 3211--3221, Oct. 2011.

\bibitem{MultirayWidebandCharact}
C. Han, A. O. Bicen, and I. F. Akyildiz,
\newblock ``Multi-ray channel modeling and wideband characterization for wireless communications in the terahertz band,''
\newblock {\em IEEE Trans. Wireless Commun.}, vol. 14, no. 5, pp. 2402--2412, May 2015.


\bibitem{StocModelTWC}
S. Priebe and T. Kurner,
\newblock ``Stochastic modeling of THz indoor radio channels,''
\newblock {\em IEEE Trans. Wireless Commun.}, vol. 12, no. 9, pp. 4445--4455, Sep. 2013.

\bibitem{StatisticalModelJSAC}
A. Saleh and R. Valenzuela,
\newblock ``A statistical model for indoor multipath propagation,''
\newblock {\em IEEE J. Sel. Areas Commun.}, vol. 5, no. 2, pp. 128--137, May 1987.

\bibitem{AoAAoDToAConf}
S. Priebe, M. Jacob, and T. Kurner,
\newblock ``AoA, AoD and ToA characteristics of scattered multipath clusters for THz indoor channel modeling,''
\newblock in {\em Proc. Eur. Conf. Antennas Propag.}, Rome, Italy, Apr. 2011, pp. 188--196.

\bibitem{song2018frequency}
S. Song, D. Liu, and F. Wang,
\newblock ``A frequency offset estimation algorithm based on under-sampling for THz communication,''
\newblock {\em in 10th ICCSN}, Chengdu, China, July 2018, 215--220.


\bibitem{SINRinSVchannelsJSAC}
W.-D. Wu, C.-C. Lee, C.-H. Wang, and C.-C. Chao,
\newblock ``Signal-to-interference-plus-noise ratio analysis for direct-sequence ultra-wideband systems in generalized Saleh-Valenzuela channels,''
\newblock {\em IEEE J. Sel. Topics Signal Process.}, vol. 1, no. 3, pp. 483--497, Oct. 2007.


\bibitem{PewithCFO}
K. Sathananthan and C. Tellambura,
\newblock ``Probability of error calculation of OFDM systems with frequency offset,''
\newblock {\em IEEE Trans. Commun.}, vol. 49, no. 11, pp. 1884--1888, Nov. 2001.

\bibitem{gao2018low}
X. Gao, L. Dai, and A. M. Sayeed,
\newblock ``Low RF-complexity technologies to enable millimeter-wave MIMO with large antenna array for 5G wireless communications,''
\newblock {\em IEEE Commun. Mag.}, vol. 56, no. 4, pp. 211-217, Apr. 2018.

\bibitem{ayach2014spatially}
O. El Ayach, S. Rajagopal, S. Abu-Surra, Z. Pi, and R. W. Heath,
\newblock ``Spatially sparse precoding in millimeter wave MIMO systems,''
\newblock {\em IEEE Trans. Wireless Commun.}, vol. 13, no. 3, pp. 1499--1513, Mar. 2014.

%\bibitem{rusu2016low}
%C. Rusu, R. Mendez-Rial, N. Gonz{\'a}lez-Prelcic, and R. W. Heath,
%\newblock ``Low complexity hybrid precoding strategies for millimeter wave communication systems,''
%\newblock {\em IEEE Trans. Wireless Commun.}, vol. 15, no. 12, pp. 8380--8393, Dec. 2016.

\bibitem{GaoBeamSelection}
X. Gao, L. Dai, Z. Chen, Z. Wang, and Z. Zhang,
\newblock ``Near-optimal beam selection for beamspace mmWave massive MIMO systems,''
\newblock {\em IEEE Commun. Lett.}, vol. 20, no. 5, pp. 1054--1057, May 2016.

%\bibitem{tsinos2017energy}
%C. G. Tsinos, S. Maleki, S. Chatzinotas, and O. Bj{\"o}rn,
%\newblock ``On the energy-efficiency of hybrid analog--digital transceivers for single-and multi-carrier large antenna array systems,''
%\newblock {\em IEEE J. Select. Areas Commun.}, vol. 35, no. 9, pp. 1980--1995, Sep. 2017.

%\bibitem{BradyBeamspaceMIMO}
%J. Brady, N. Behdad, and A. Sayeed,
%\newblock ``Beamspace MIMO for millimeterwave communications: System architecture, modeling, analysis, and measurements,''
%\newblock {\em IEEE Trans. Antennas Propag.}, vol. 61, no. 7, pp. 3814--3827, July 2013.

\bibitem{AALimitedFeedback}
A. Alkhateeb, G. Leus, and R. W. Heath,
\newblock ``Limited feedback hybrid precoding for multi-user millimeter wave systems,''
\newblock {\em IEEE Trans. Wireless Commun.}, vol. 14, no. 11, pp. 6481--6494, Nov. 2015.

\bibitem{Sohrabi2016hybrid}
F. Sohrabi and W. Yu,
\newblock ``Hybrid digital and analog beamforming design for large-scale antenna arrays,''
\newblock {\em IEEE J. Sel. Topics Signal Process.}, vol. 10, no. 3, pp. 501–-513, Apr. 2016.

\bibitem{BWangNOMAmmWave}
B. Wang, L. Dai, Z. Wang, N. Ge, and S. Zhou,
\newblock ``Spectrum and energy efficient beamspace MIMO-NOMA for millimeter-wave communications using lens antenna array,''
\newblock {\em IEEE J. Select. Areas Commun.}, vol. 35, no. 10, pp. 2370--2382, Oct. 2017.


\bibitem{SpatialCCTSP}
S. Park, J. Park, A. Yazdan, and R. W. Heath,
\newblock ``Exploiting spatial channel covariance for hybrid precoding in massive MIMO systems,''
\newblock {\em IEEE Trans. Signal Process.}, vol. 65, no. 14, pp. 3818--3832, July 2017.

\bibitem{yuan2019low}
H. Yuan, J. An, N. Yang, K. Yang, and T. Q. Duong,
\newblock ``Low complexity hybrid precoding for multiuser millimeter wave systems over frequency selective channels,''
\newblock {\em IEEE Trans. Veh. Technol.}, vol. 68, no. 1, pp. 983–-987, Jan. 2019.

\bibitem{liang2014low}
L. Liang, W. Xu, and X. Dong,
\newblock ``Low-complexity hybrid precoding in massive multiuser MIMO systems,''
\newblock {\em IEEE Wireless Commun. Lett.}, vol. 3, no. 6, pp. 653–-656, Dec. 2014.

%\bibitem{shen2019channel}
%W. Shen, L. Dai, J. An, P. Fan, and R. W. Heath,
%\newblock ``Channel estimation for orthogonal time frequency space (OTFS) massive MIMO,''
%\newblock {\em IEEE Trans. Signal Process.}, vol. 67, no. 16, pp. 4204--4217, Aug. 2019.

\bibitem{du2012probabilistic}
H. Du and P.-J. Chung,
\newblock ``A probabilistic approach for robust leakage-based MU-MIMO downlink beamforming with imperfect channel state information,''
\newblock {\em IEEE Trans. Wireless Commun.}, vol. 11, no. 3, pp. 1239--1247, Mar. 2012.

\bibitem{J2018frequency}
J. Rodriguez-Fernandez, N. Gonzalez-Prelcic, K. Venugopal, and R. W. Heath,
\newblock ``Frequency-domain compressive channel estimation for frequency-selective hybrid millimeter wave MIMO systems,''
\newblock {\em IEEE Trans. Wireless Commun.}, vol. 17, no. 5, pp. 2946--2960, May 2018.

%\bibitem{khalid2014robust}
%F. Khalid and J. Speidel,
%\newblock ``Robust hybrid precoding for multiuser MIMO wireless communication systems,''
%\newblock {\em IEEE Trans. Wireless Commun.}, vol. 13, no. 6, pp. 3353--3363, June 2014.

\bibitem{grantCVX}
M. Grant and S. Boyd,
\newblock {\em CVX: MATLAB Software for Disciplined Convex Programming, Version 2.1.} [Online]. Available: https://cvxr.com/cvx








\end{thebibliography}
\end{document}